\documentclass[aps,twocolumn,pra,superscriptaddress,amsmath,showpacs,tightenlines]{revtex4-1}
\usepackage{amssymb}
\usepackage{amsmath}
\usepackage{graphicx}
\usepackage{subfigure}
\usepackage{natbib}
\usepackage{epsfig}
\usepackage{amsfonts}
\usepackage{mathrsfs}
\usepackage{CJK}
\usepackage{xcolor}
\usepackage[toc,page,title,titletoc,header]{appendix}

\begin{document}
\title{Quantum interference and controllable magic cavity QED  via a giant atom in coupled resonator waveguide}
\author{Xiaojun Zhang}
\affiliation{Center for Quantum Sciences and School of Physics, Northeast Normal University,
Changchun 130024, China}
\author{Chengge Liu}
\affiliation{Department of Physics, Applied Optics Beijing Area Major Laboratory, Beijing Normal University, Beijing 100875, China}
\author{Zhirui Gong}
\affiliation{College of Physics and Optoelectronic Engineering, Shenzhen University, Shenzhen, 518060, China}
\author{Zhihai Wang}
\email{wangzh761@nenu.edu.cn}
\affiliation{Center for Quantum Sciences and School of Physics, Northeast Normal University,
Changchun 130024, China}

\begin{abstract}
We study the Markovian and Non-Markovian dynamics in a giant atom system which couples to a coupled resonator waveguide (CRW) via two distant sites. Under certain conditions, we find that the giant atom population can exhibit an oscillating behavior and the photon can be trapped in the giant atom regime. These phenomena are induced by the interference effect among the bound states both in and outside the continuum, which is peculiar for two sites atom-CRW coupling. As an application of the photon trapping, we theoretically propose a magic cavity model where the giant atom serve as either a perfect or leaky cavity, depending on the distance between the coupling sites. The controllability of the magic cavity from perfect to leaky one can not be realized in the traditional cavity or circuit QED setup. The predicted effects can be probed in state-of-the-art waveguide QED experiments and provide a striking example of how the different kinds of bound states modify the dynamics of quantum open system in a structured environment.
\end{abstract}
%\pacs{42.50.Pq, 03.67.Lx, 42.50.Dv}
\maketitle
\section{Introduction}

Ever since the pioneering work which couples the transmon qubits to the propagating surface acoustic wave~\cite{S1,S2}, the giant atom (GA) which can also be realized by superconduting qubit~\cite{sq1,sq2,sq3} and magnon spin ensemble~\cite{mq}, has attracted considerable attentions in the context of waveguide QED. Beyond the dipole approximation in the conventional quantum optics treatment, the GA couples to the waveguide via more than one connecting points~\cite{g5}. The nonlocal light-matter interaction in GA gives rise lots of interesting physical effects, such as frequency dependent relaxation~\cite{ar2022,Lambpra14}, decoherence-free interaction~\cite{AF2018,AC2020}, chiral photonic population~\cite{XW2021,XW2022,AS2021,NL2022}, non-Markovian oscillating~\cite{DZ2021,Guoprr20,SG2020}, phase controlled entanglement~\cite{Jieqiao,PRL2023},  to just name a few. In these works, the non-negligible time and phase accumulation as the photon propagates between/amomg the atom-waveguide coupling points play predominant roles.

The GA in most of the previous studies is usually assumed to couple to the waveguide with linear dispersion relation~\cite{xian1,xian2,xian4}, in which the group velocity of the photon is independent of the wave vector. However, the discrete site waveguide via tight binding interaction can be constructed by photonic crystals or superconducting quantum circuits with the nowadays available technologies~\cite{c1,c2,c3,c4}. Such waveguide supports {quasi continual} band structure with cosine
dispersion relation. Therefore, a natural question arises:
how the GA behaves in a structured environment which is composed by the coupled resonator waveguide (CRW). In this paper, we address this question by analyzing a system where a two-level GA couples a CRW via two coupling points. Appropriately choosing the distance between the two coupling points, we can achieve the atomic oscillation beyond the Markovian dynamics, in which case, the photon is trapped inside the GA regime via quantum interference effect. We find that this behavior is led by the interference between the bound states in the continuum (BIC)~\cite{BIC1,BIC2,BIC3,BIC4,BIC5} and outside of the continuum (BOC)~\cite{BOC1,BOC2,BOC3,BOC4}. It is therefore dramatically different from the mechanism of oscillating bound state in linear waveguide and the CRW in Ref.~\cite{limpra}, where only the BICs involve with some harsh terms. Thanks to the BIC-BOC interference mechanism in our two sites coupling scheme, it becomes much easier to implement compared with the multiple coupling points scheme in Ref.~\cite{limpra}.

As an application of the photon trapping by GA, we propose an effective magic cavity QED setup by introducing another auxiliary conventional small atom, which locates between the coupling points between the GA and CRW. Here, the GA serves as a controllable magic cavity and the small atom plays as the emitter. Different from the magic cavity formed by two small atoms in linear or nonlinear waveguide~\cite{mc1,mc2,mc3}, our effective magic cavity can be either a perfect or leaky one, depending on the distance between the atom-waveguide coupling points, due to its size dependent decay rate. We further show the Rabi splitting and oscillation in the perfect cavity and dark state, which is a BIC in the leaky cavity limit, respectively.

\section{Single giant atom}
\label{single}

\begin{figure}
  \centering
  \includegraphics[width=1\columnwidth]{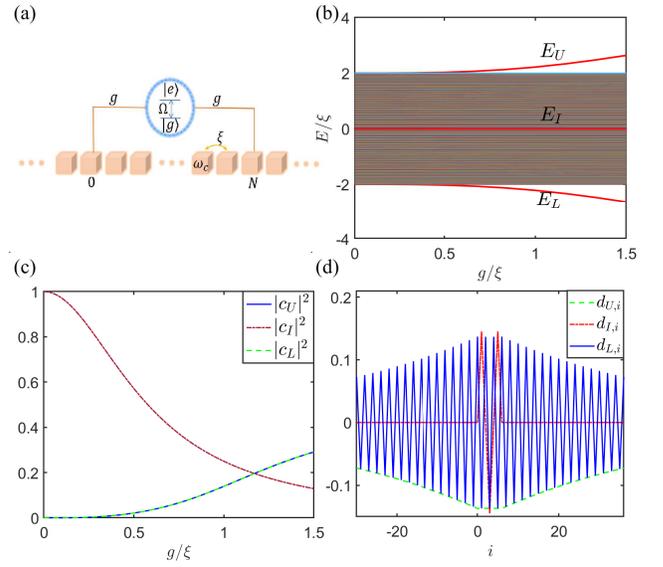}
  \caption{(a) Sketch of the waveguide QED setup, where a GA is coupled to a CRW with via two sites with label $0$ and $N$. (b) The corresponding energy diagram in the single excitation subspace. (c) Atomic population and (d) photonic amplitudes for the bound state in and outside the continual band. The parameters are set as $\Omega=\omega_c=0$, $N=6$ in (b-d) and {$g=0.15\xi$} in (d). }\label{scheme1}
\end{figure}

As schematically shown in Fig.~\ref{scheme1}(a), we begin with the model that a single GA couples to a CRW via two coupling sites, which are labeled by $0$ and $N$, respectively. The CRW is modeled by the tight-binding Hamiltonian (hereafter $\hbar=1$)
\begin{equation}
H_{C}=\omega_c \sum_{j}a_{j}^{\dagger}a_{j}-\xi \sum_{j}(a_{j+1}^{\dagger}a_{j}+a_{j}^{\dagger}a_{j+1})
\end{equation}
where $a_j$ is the photon annihilation operator of the $j$th resonator, $\omega_c$ and $4\xi$ are the central frequency and the total width of the
continuum, respectively. The Hamiltonian of the whole system including the GA and the CRW is
\begin{equation}
H_s=H_{C}+\Omega \sigma_+\sigma_-+g[(a_{0}^{\dagger}+a_{N}^{\dagger})\sigma_- +(a_{0}+a_{N})\sigma_+ ]
\end{equation}
where $\Omega$ is the transition frequency of the GA between its ground state $|g\rangle$ and excited state $|e\rangle$, $\sigma_+=(\sigma_-)^\dagger=|e\rangle\langle g|$ is the {raising} operator of the GA, and $g$ is the coupling strength between the GA and the resonator in the waveguide.

{Now, we resort to the Fourier transformation $a^{\dagger}_{k}=\sum_{j}e^{-ikj}a^{\dagger}_{j}/\sqrt{N_c}$
with {$N_c\to\infty$} being the length of the CRW, the Hamiltonian in the momentum space yields
\begin{equation}
H_s=\Omega \sigma_+\sigma_-+\sum_{k}\omega_k a_{k}^{\dagger}a_{k}
+\sum_{k}[ g_k a_{k}^{\dagger}\sigma_{-}+\rm H.c.].\label{mo}
\end{equation}
Here, the dispersion relation of the waveguide satisfies $\omega_k=\omega_c-2\xi\cos k$, $g_k=g(1+e^{ikN})/\sqrt{N_c}$ and the coupling strength $g$ is considered to be real. In what follows, we will consider that the giant atom is resonant with the bare resonator in the waveguide and set their frequency as zero, that is, $\Omega=\omega_c=0$. }

In what follows, we are interested in a scenario where $\Omega$ lies in the center of the continuum, such that the waveguide supplies an effective structured environment for the GA. As shown in Fig.~\ref{scheme1}(b), where we illustrate the energy spectrum of the whole GA-CRW coupled system in the single excitation subspace as a function of the coupling strength $g$ by setting $N=6$.  {We observe a continual band  in the frequency regime between  $-2\xi$ and $2\xi$, which is same to that without the giant atom. Moreover, the giant atom gives birth to three exotic states ($E_U$, $E_L$ and $E_I$), which are denoted by the red solid lines in Fig.~\ref{scheme1}(b). To study them in detail, we assume the single excitation wave function in the momentum state as $|E_\alpha\rangle=(\sum_k d_{\alpha,k}a_k^\dagger+c_{\alpha}\sigma_+)|g, \rm {vac}\rangle$, the eigen energy $E$ then satisfies the transcendental equation (see Appendix A  for the detailed derivations) }

{\begin{equation}
E=\frac{g^2}{\pi}\int_{-\pi}^{\pi} dk\frac{1+\cos kN}{E+2\xi\cos k}.
\label{equationx}
\end{equation}}

{In the regime of $|E|>2\xi$, we find a pair of solutions, which locate outside the continuum~\cite{OBC1,Peter16pra} and are denoted by $E_U $ and $E_L$ respectively. These two energies come from the break down of the translational symmetry, which is induced by the GA-CRW coupling. As the increase of coupling strength, they gradually depart from the upper and lower boundary of the continuum, and therefore we name them as BOC. One should note that, these BOCs are not unique for the GA. For example, when a small atom couples to the CRW, we can still observe the BOCs~\cite{Peter16pra}. These two BOCs are actually the atom-photon hybrid state. For the atom partner, as show in Fig.~\ref{scheme1}(c), the atomic population satisfies $|c_{U}|=|c_{L}|$ and increases with the GA-CRW coupling strength. This trend is  also similar to that in the small atom setup~{\cite{Peter16pra}}.  For the photonic partner, we show that it is centralized at the at two legs of the GA and exponentially decay along two directions, this is why we named them BOC. Meanwhile, is satisfies the symmetry relation $d_{U,j}=(-1)^{j+1}d_{L,j}$, where $d_{\alpha,j}=\sum_k b_k e^{-ikj}/\sqrt{N_c}$ is the photonic excitation amplitudes in the $j$th site for the state $|E_\alpha\rangle$.}

{Besides, we also unexpectedly find that $E=0$ is a solution to Eq.~(\ref{equationx}), which is denoted by $E_I$ in Fig.~\ref{scheme1}(b). Since it locates inside the continuum, and hybrids the atomic and photonic excitation as shown in Figs.~\ref{scheme1}(c) and (d), we name it BIC~\cite{IBC1,SL}. In Fig.~1(c), we sketch the atomic population $|c_{\alpha}|^2$ for the BOCs and BIC. For the BIC, the single excitation is mainly on the atom in the weak coupling regime, and continually decreases as the coupling strength increases. For the other eigen states except for these three bound states, there is nearly no atomic excitation.   Meanwhile, different from the BOCs,  the photonic excitation probability is only valued at the $1$st, the $3$rd, and the $5$th lattices which are between the two legs of the GA for BIC. The photonic amplitudes satisfy { $d_{I,1}=-d_{I,3}=d_{I,5}$} (see the red curve in Fig.~\ref{scheme1}(d)) and more discussions about the BIC are given in Appendix A. }

{The above results show that the BIC exists in the single two-leg giant atom system when $\Omega=\omega_c, N=6$. We also find that the BIC is always present when $\sum_{1}^{M}\exp(iKn_j)=0$ in a single giant atom with more than two legs. Here $K$ satisfies $\Omega=\omega_c-2\xi\cos K$, $n_j$ is the position of the $jth$ coupling point in the coupled resonator waveguide and $M$ is the number of the total coupling points between the giant atom and the CRW, this result agrees with that given in Ref.~\cite{limpra}}.

\subsection{{Markovian dynamics}}

In Fig.~\ref{decay}(a) and (b), we show the evolution of the excited state population $P_e(t)=\langle |e\rangle\langle e|\rangle$ for the initially excited GA. Under the Markov approximation, the master equation is obtained as
(see Appendix B and Ref.~\cite{wei} for the detailed derivation)
\begin{equation}
\dot{\rho}=-i\Omega[|e\rangle\langle e|,\rho]+(A+A^{*})\sigma^{-}\rho\sigma^{+}
-A\sigma^{+}\sigma^{-}\rho-A^{*}\rho\sigma^{+}\sigma^{-}
\end{equation}
where
\begin{equation}
A=\frac{g^2}{\xi}(1+e^{i\pi N/2}).
\end{equation}
In the case of $N=4$, we show an exponential decay in Fig.~\ref{decay}(a) for the atomic excitation. It shows that, the results based on the Markovian approximation agree with the numerical results well. However, the Makorvian approximation is not valid for $N=6$, and the numerical result in Fig.~\ref{decay}(b) yields a small decay initially and a nearly steady oscillation after a long time evolution. In Fig.~\ref{decay} (c) and (d), we furthermore numerically illustrate the evolution of the photonic distribution under the same initial state. For $N=4$, the photon emitted by the GA occupies in the resonator labelled by $j=2$ and gradually diffuses to the whole waveguide as shown in Fig.~\ref{decay} (c). Therefore, the waveguide, which acts as the memoryless environment, leads to an exponential atomic decay, being similar to that under the Markovian approximation as shown in Fig.~\ref{decay}(a). For $N=6$, the photon is trapped in the resonators between the two legs of GA with site number label $j=1,3,5$, leading an obvious non-Markovian atomic evolution. The above atomic decay (oscillation) phenomena also occurs for $N=4m$ ($N=4m+2$) for arbitrary integer $m$ {when $g\ll\xi$~\cite{wei}.}

\subsection{{Non-Markovian dynamics}}

\begin{figure}
  \centering
  \includegraphics[width=0.5\columnwidth]{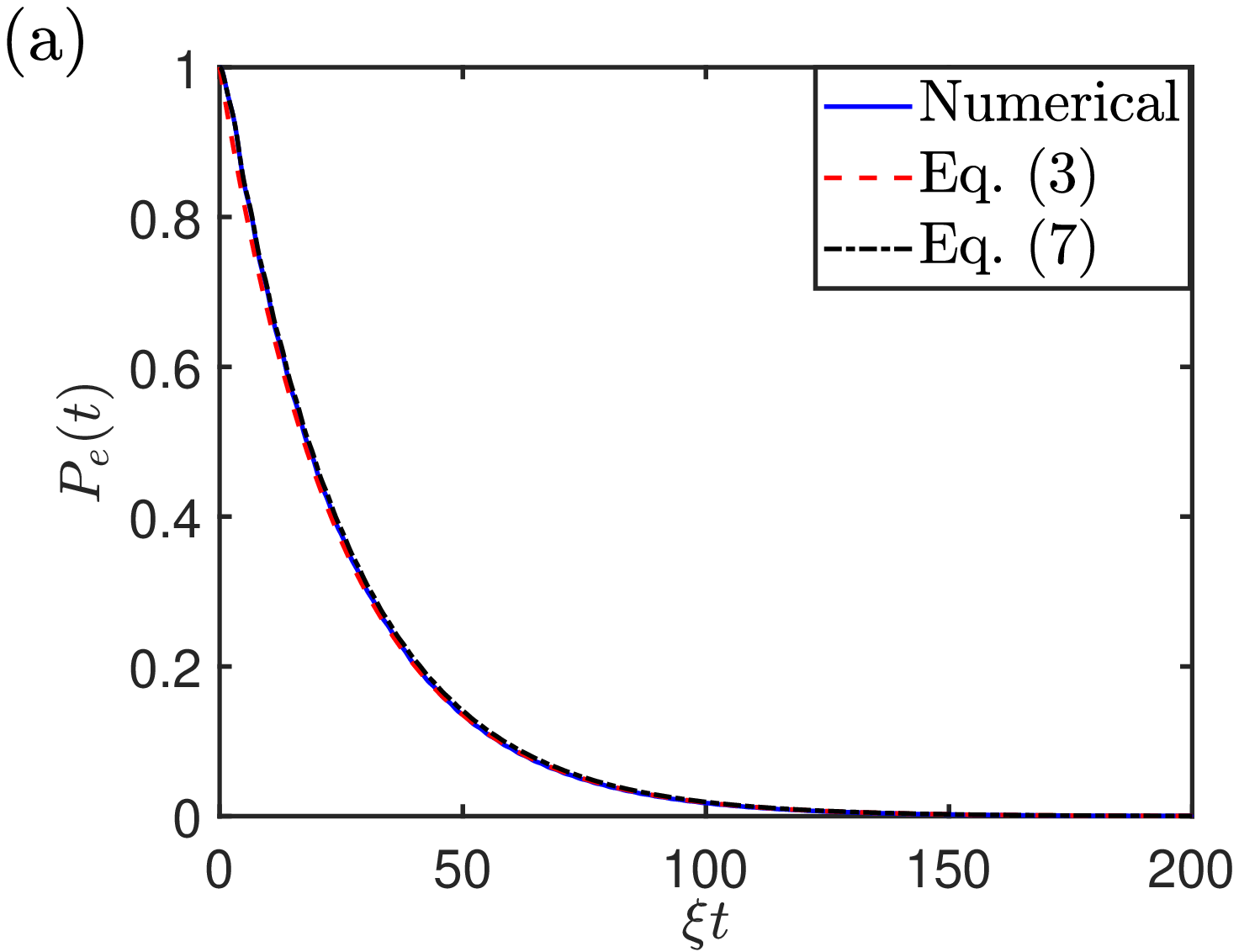}\includegraphics[width=0.5\columnwidth]{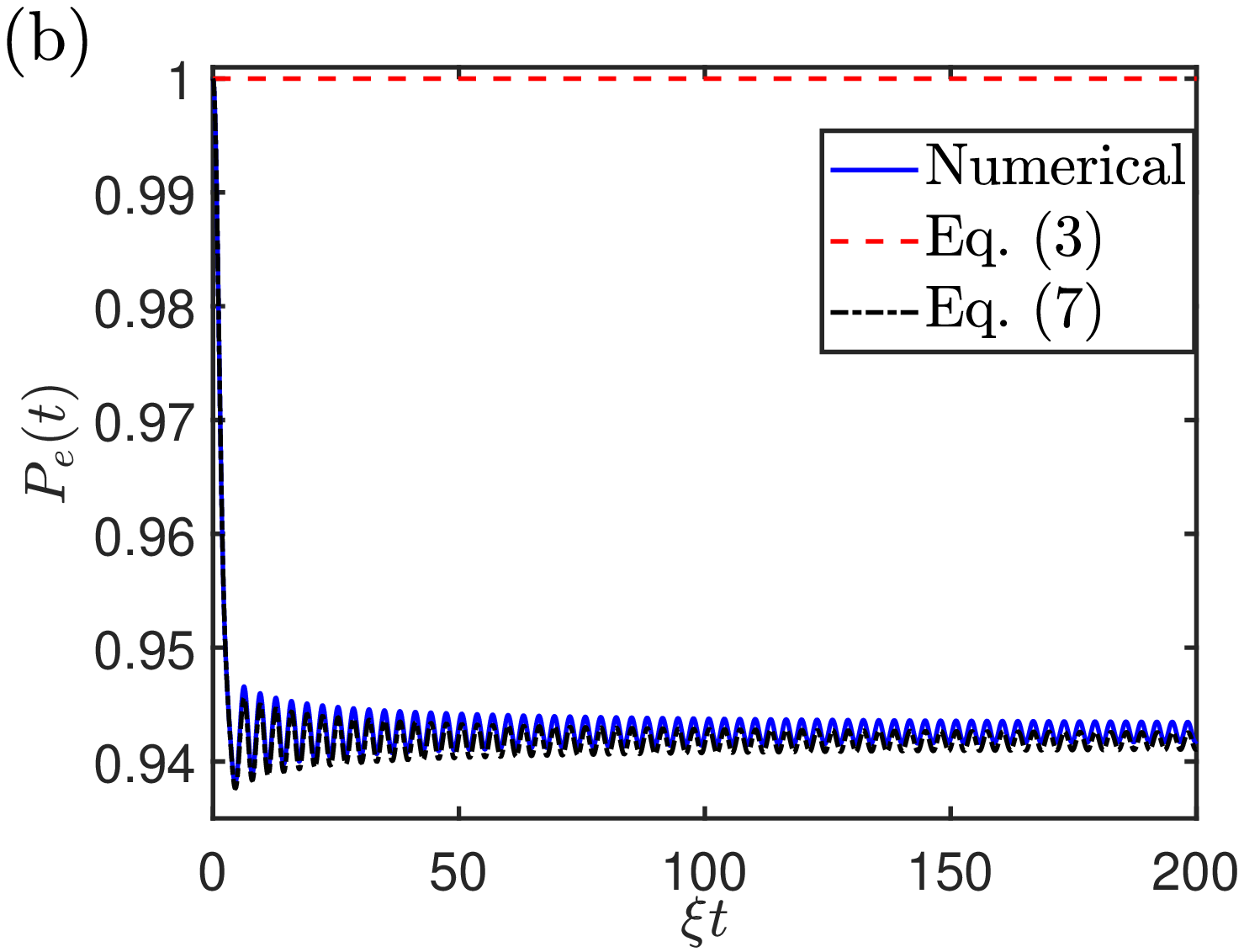}\\
  \includegraphics[width=0.5\columnwidth]{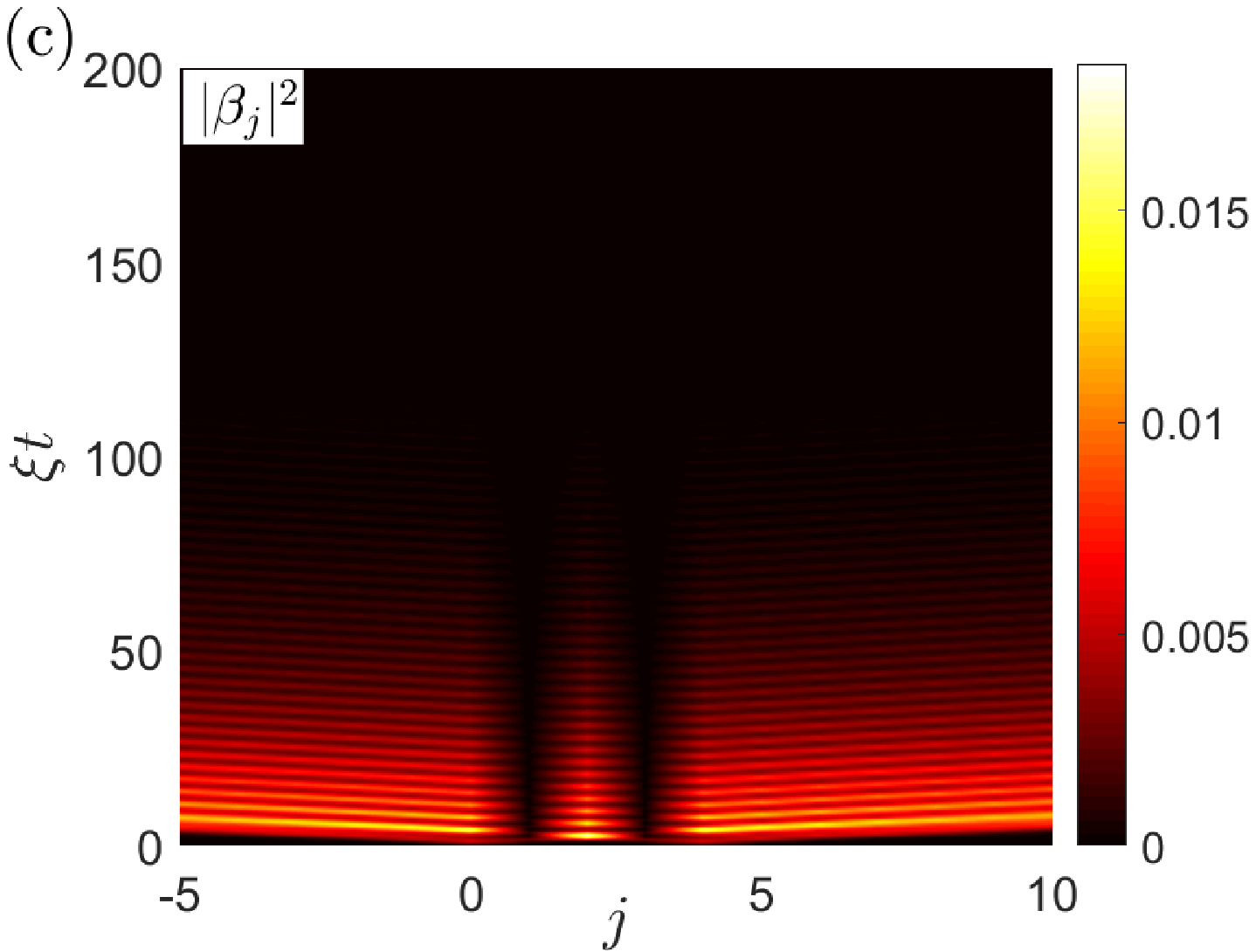}\includegraphics[width=0.5\columnwidth]{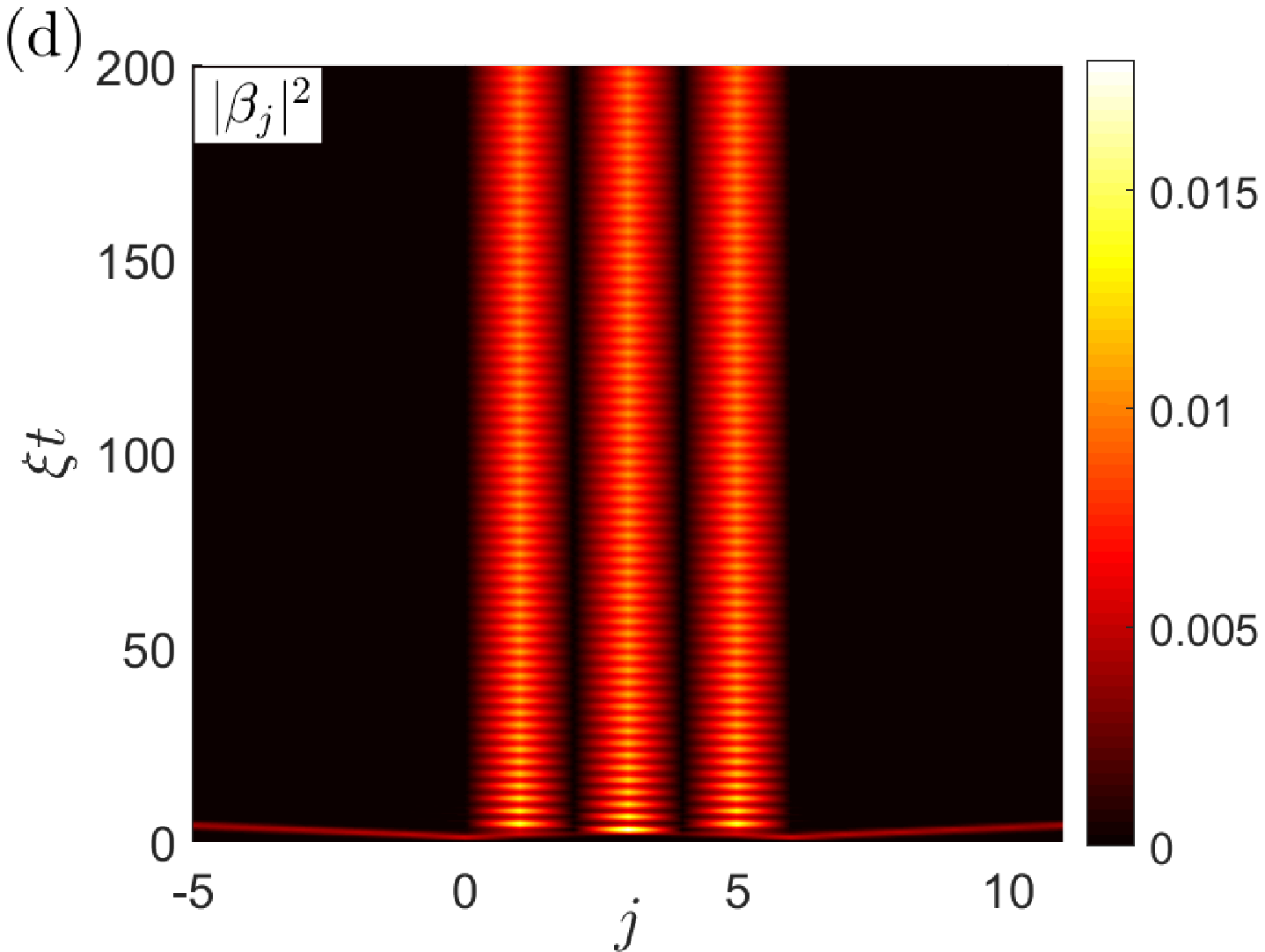}\\
  \includegraphics[width=0.5\columnwidth]{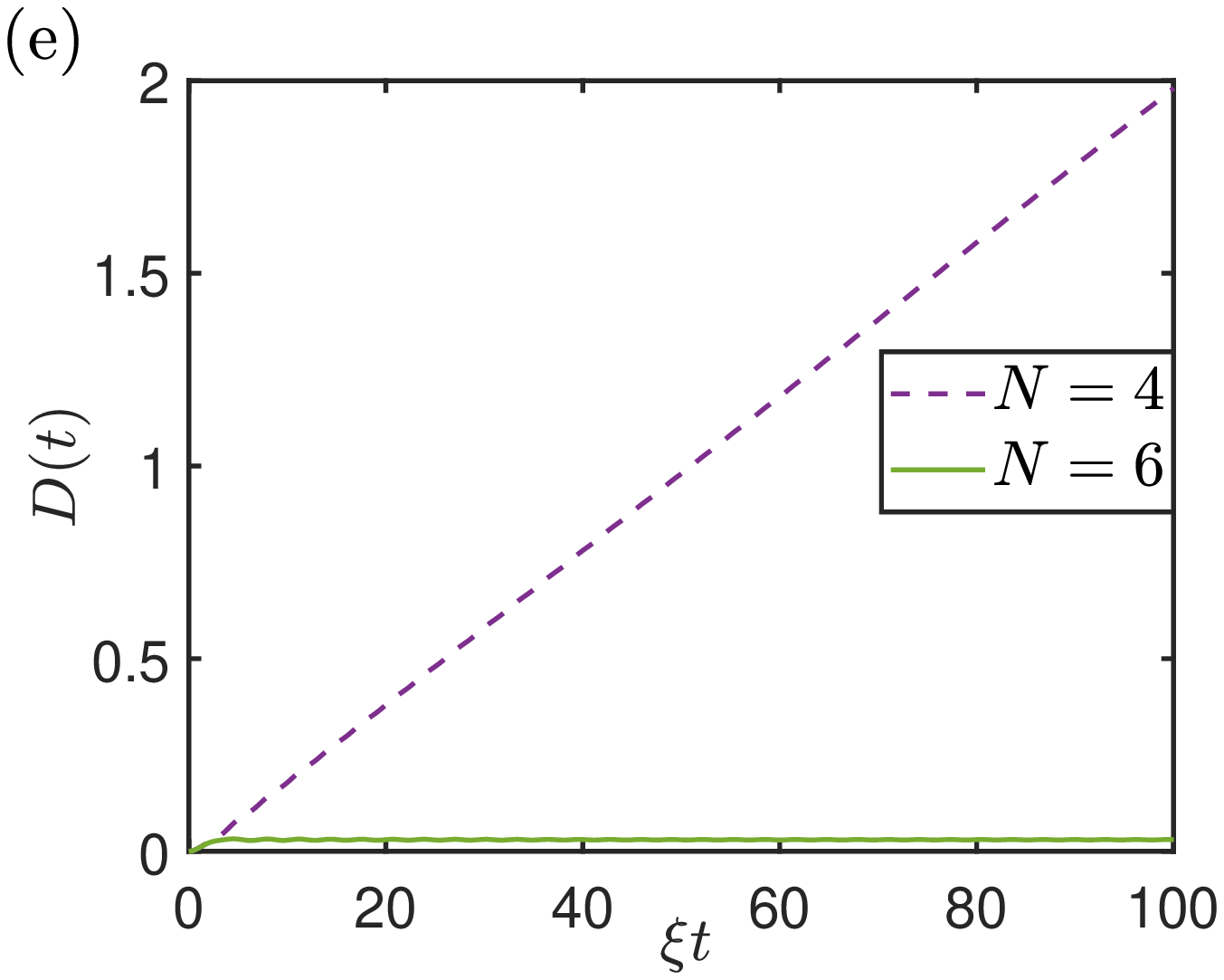}\includegraphics[width=0.5\columnwidth]{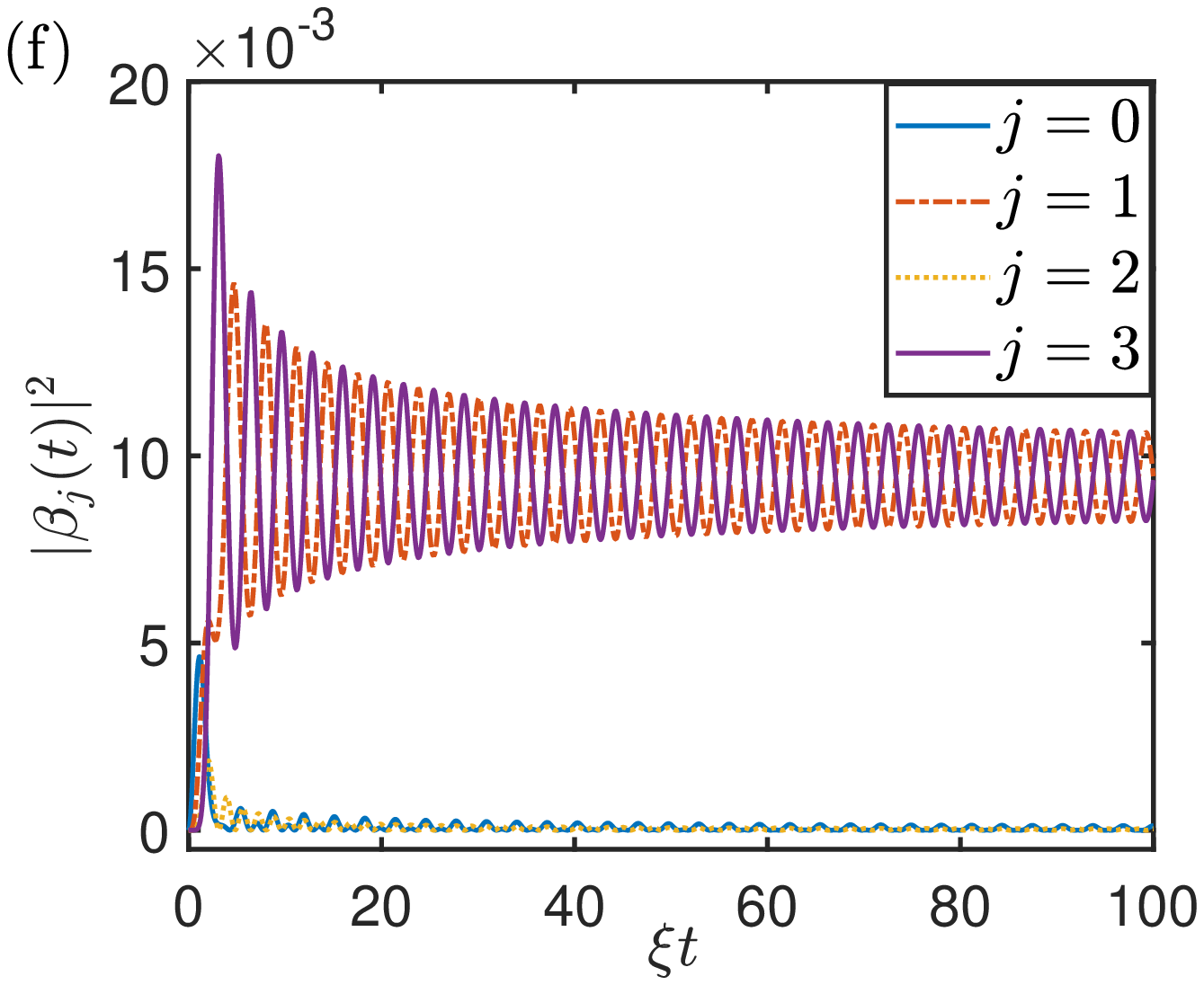}
  \caption{Evolution of the atomic excited state {(a,b)} and photonic population {(c,d)}. We set $N=4$ in (a,c) and $N=6$ in (b,d). {(e) Time evolution of $D(t)$ in Eq.~(\ref{exp}). (f) Photonic population inside of regime of the giant atom for $N=6$. The parameters are set as $\Omega=\omega_c,g=0.1\xi$.}}\label{decay}
\end{figure}

To understand how the quantum interference effect leads to these dramatically different dynamical behaviors in Fig.~\ref{decay} (a) and (b). {Base on the Eq.~(\ref{mo}),
 we write the single excitation wave function as}
 \begin{equation}
\left|\psi(t)\right\rangle=e^{-i\Omega t}[ \alpha (t)\sigma_+|G\rangle+\sum_{k}\beta_{k}(t)a_{k}^{\dagger}\left|G\right\rangle],
\end{equation}
and perform some detailed calculations under the Weisskopf-Wigner approximation as shown in Appendix C, the atomic excitation amplitude can be obtained as
\begin{equation}
{\alpha(t)=e^{-D(t)},}\label{exp}
\end{equation}
{where
$D(t)=-2g^2\int_{0}^{t}dt_1 \int_{0}^{t_1}d\tau G(\tau)$, and $G(\tau)=J_0(2\xi\tau)+i^N J_N(2\xi\tau)$.
In Fig.~2(a,b), we give the atomic dynamics based on Eq.~(\ref{exp}).
When $N = 4$, the results of Non-Markovian approximation based on Eq.~(\ref{exp}) agree with the numerical simulation well. However, for $N = 6$, the Markovian approximation breaks down and Eq.~(\ref{exp}) predicts a valid result comparing to the numerical calculations.}

Furthermore, the photonic population in the real space is
\begin{eqnarray}
\beta_j=\frac{1}{\sqrt{N_c}}\sum_{k}\beta_k e^{-ikj}
=-ig\int_{0}^{t}d \tau \alpha(t-\tau)F_j(\tau),\label{beta}
\end{eqnarray}
where the quantum interference effect can be extracted from the function {$G(\tau)=J_0(2\xi\tau)+i^N J_N(2\xi\tau)$} and $F_j(\tau)=i^{j} J_{j} (2\xi\tau)+i^{j-N}J_{j-N}(2\xi\tau)$, with $J_{m}$ being the $m$th order Bessel function. First, for the case of $N=4$, the constructive interference with {$G(\tau)=J_0+ J_{4}$} leads to a fast atomic decay. As for the photon distribution, the constructive interference $F_2(\tau)=-2J_2(2\xi\tau)$ and the destructive interference
$F_1(\tau)=F_3(\tau)=i[J_1(2\xi\tau)-J_3(2\xi\tau)]$ leads to the photonic occupation in the resonator with $j=2$ at the early time as shown in Fig.~\ref{decay}(c). Second, for $N=6$, the destructive interference with $G(\tau)=J_0(2\xi\tau)- J_{6}(2\xi\tau)$ leads to the dissipation suppression and {constructive ({destructive})} interference between the two Bessel functions in $F_j$ for $j=1,3,5$ ({2,4}) leads to the striped photonic distribution as shown in Fig.~\ref{decay}(d).{In Fig.~\ref{decay}(e), we plot the curve for $D(t)$ as a function of the evolution time $t$.  We observe that $D(t)$ will obtain a relative larger positive value for $N=4$ than that of $N=6$, while the former one increases with time and the later one keeps a steady small value as the time evolution.
This fact agrees with the results in Fig.~\ref{decay} (a) and (b) in that the atom will dissipation fast for $N=4$ while keep a large excitation state population for $N=6$.  Furthermore, according to Eq.~(\ref{beta}), we give the distribution of photons in $0,1,2,3$ lattices in the waveguide when $N = 6$ as sketched in Fig.~~\ref{decay} (f). The result is completely consistent with Fig.~2(d), in which the second site has almost no photons, and the photons are concentrated in the first and third sites.}

\subsection{BIC-BOC oscillation}
\begin{figure}
  \centering
  \includegraphics[width=1\columnwidth]{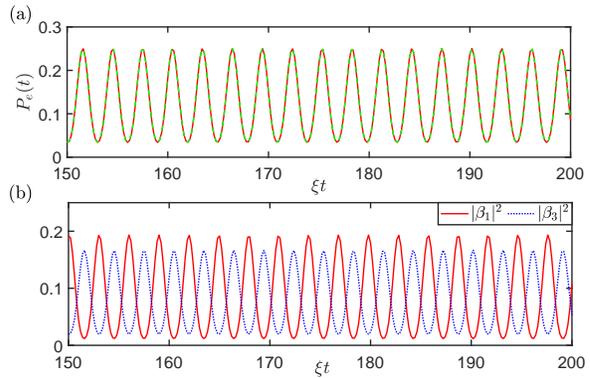}
  \caption{(a) Evolution of atomic excitation probability based on numerical calculation (red solid line) and Eq.~(\ref{BIC}) (red dashed line). (b) Evolution of photonic population $|\beta_1|^2$ and $|\beta_3|^2$. The parameters are set as $N=6,\Omega=\omega_c=0,g=0.8\xi$.}
  \label{BICf}
\end{figure}

Meanwhile, the small oscillation in Fig.~2 (b) and (d) for $N=6$ indicates that the GA coherently exchange excitation with the photon in the waveguide. The oscillation behavior can be explained by the interference between the BIC and BOCs during the time evolution. Enlightened by the atomic population for BIC and BOCs as shown in Fig.~1(c), the initial state $|\psi(0)\rangle=\sigma_+|g,{\rm vac}\rangle$ can be written as
$|\psi(0)\rangle= c_{U}|E_U\rangle+c_{L}|E_L\rangle+c_{I}|E_I\rangle+\sum_cc_k|E_k\rangle$, where
 $|E_k\rangle$ is the states in the continuum.
Then, the state of the system {at long time moment $t$} is obtained as
$|\psi(t)\rangle=e^{-iE_It}[e^{-i\delta t}c_{U}|E_U\rangle+e^{i\delta t}c_{L}|E_L\rangle+c_{I}|E_I\rangle]$ where $\delta=E_U-E_I=E_I-E_L$ is the detuning between the BIC and the BOC and the states in the continuum will play no roles to the atomic population when the evolution time is long enough.  As a result, we will obtain
\begin{eqnarray}
|\alpha(t)|^2&=&|c_I^2+2c_U^2\cos(\delta t)|^2,\label{BIC}\\
|\beta_j(t)|^2&=&{|e^{-i\delta t}c_Ud_{U,j}+e^{i\delta t}c_Ld_{L,j}+c_Id_{I,j}|^2}\label{BIC2}.
\end{eqnarray}
Therefore, the transition between BOCs and BIC induces the excitation exchange between GA and photon with the period $T=2\pi/\delta$, and we show the agreement between the results based on the above equation and those obtained numerically in Fig.~\ref{BICf} (a). To show an obvious oscillation, we here choose a larger GA-waveguide coupling strength $g=0.8\xi$, in which the Weisskopf-Wigner approximation does not work any more. In Fig.~\ref{BICf} (b), we illustrate the photonic population, which shows that the photon oscillates between the different resonators with odd labels.

\section{Magic cavity QED model}

{The above results show that the GA with appropriate size will effectively prevent the photon in the waveguide between the coupling points from escaping, which is similar with two atoms setup~\cite{mc1,mc2,mc3}. It has the same effect as the traditional optical cavity, that is, trapping photons. Followed by the two atom setup with linear waveguide~\cite{mc3} or CRW~\cite{mc2}, we name this giant atom as a magic cavity.} One of the promising applications is to construct a controllable cavity-QED setup, by effectively coupling a conventional small atom to the magic cavity. To this end, we introduce an auxiliary small two-level atom, which has the same transition frequency $\Omega$ with the GA and is located in the $M$th ($0<M<N$) resonator. As shown in Fig.~4(a), the Hamiltonian of the magic cavity QED model is written as
\begin{equation}
H_{\rm QED}=H_s+\Omega \tau_+\tau_-+g_s (\tau_+a_M+a_M^\dagger\tau_-),
\end{equation}
where $\tau_{\pm}$ is the Pauli operator for the auxiliary small atom, and real $g_s$ is its coupling strength to the $M$th resonator in the CRW.

Taking the CRW as the structured environment and performing the Markovian  approximation, the dynamics of the magic cavity QED mode is governed by the master equation~(see Appendix A)
\begin{eqnarray}
\frac{d}{dt}\rho&=&-i[H_{\rm eff},\rho]+\gamma_g D_{[\sigma_+,\sigma_-]}\rho
+\gamma_s D_{[\tau_+,\tau_-]}\rho\nonumber \\
&&+\gamma_I \left(D_{[\tau_+,\sigma_-]}\rho+D_{[\sigma_+,\tau_-]}\rho\right),\label{master}
\end{eqnarray}
where $D_{[O_1,O_2]}\rho=2O_2\rho O_1-\rho O_1O_2-O_1O_2\rho$. Here, $\gamma_g={\rm Re} A, A=g^2(1+i^N)/\xi$ and $\gamma_s=g_s^2/(2\xi)$ are the individual decay rate of the giant and small atoms, respectively. $\gamma_I={\rm Re} B, B=gg_s(i^M+i^{(N-M)})/(2\xi)$ is their collective decay rate, which is induced by the common CRW environment. Taking the classical driving to the GA into consideration, the effective Hamiltonian in the rotating frame is expressed as
\begin{eqnarray}
H_{\rm eff}&=&(\Delta+\delta_g)\sigma_+\sigma_-+\Delta\tau_+\tau_-\nonumber \\
&&+g_I(\sigma_+\tau_-+\tau_+\sigma_-)+\eta(\sigma_++\sigma_-),
\end{eqnarray}
where $\Delta$ is the detuning between the atoms and the driving field, and $\eta$ is the driving strength. $\delta_g={\rm Im}A$ and $g_I={\rm Im} B$ are the CRW induced Lamb shift for the GA and effective coupling strength between the two atoms, respectively.

\begin{figure}
  \centering
  \includegraphics[width=0.5\columnwidth]{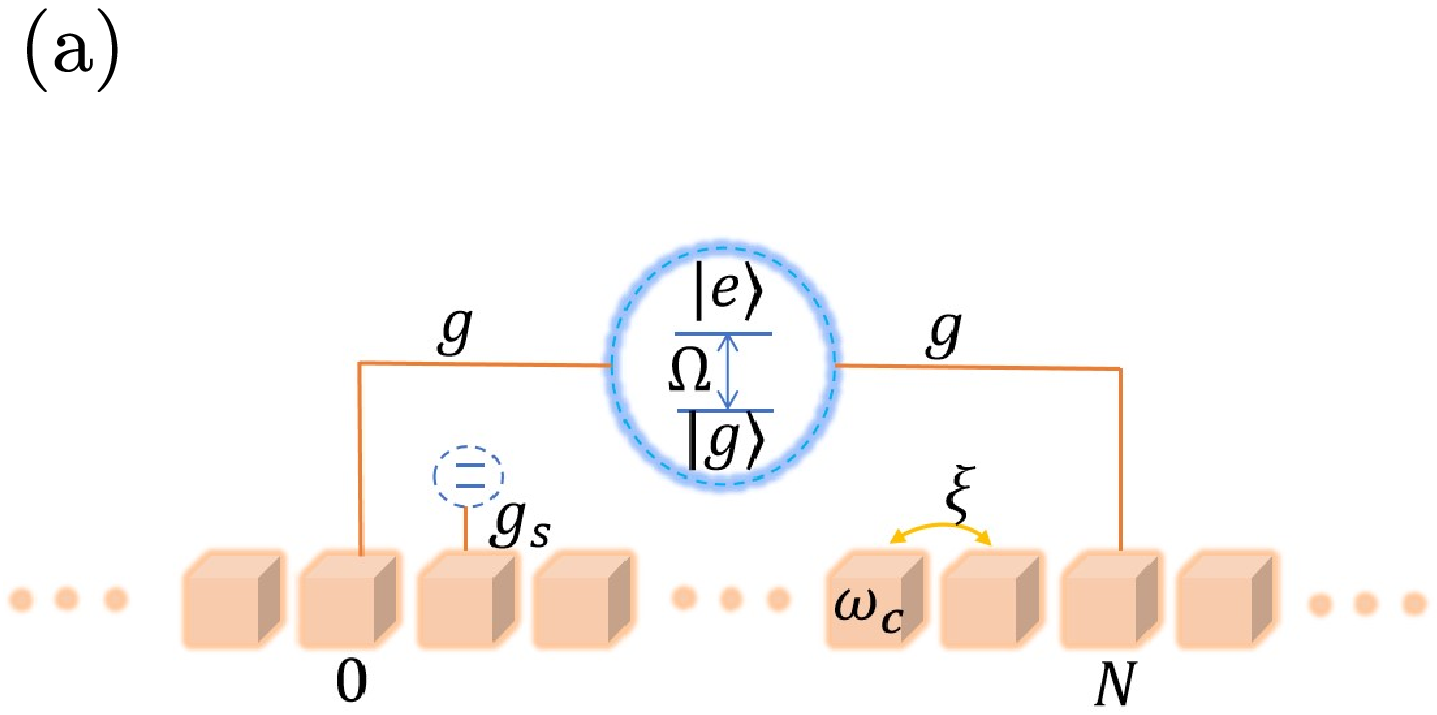}\includegraphics[width=0.5\columnwidth]{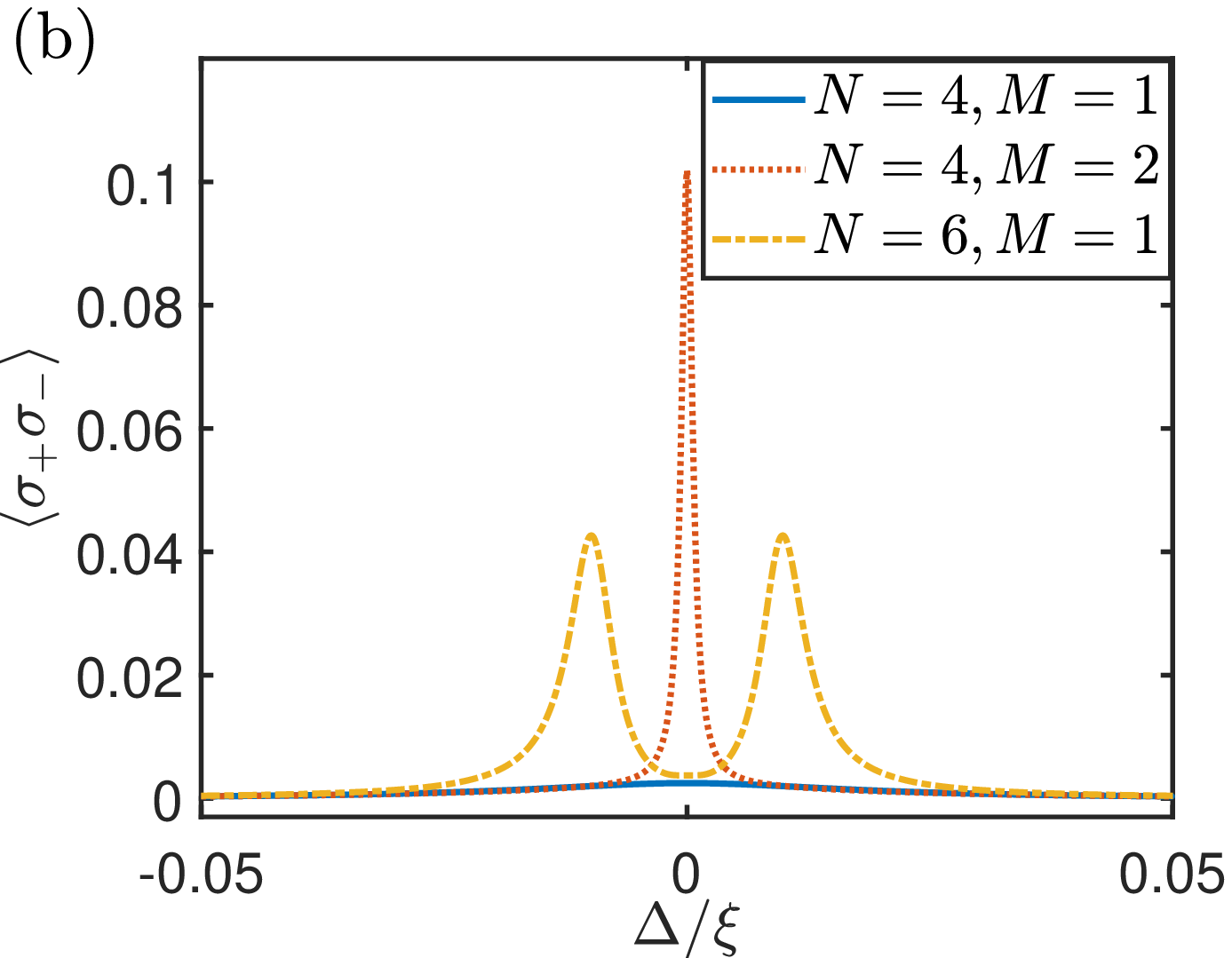}\\
  \includegraphics[width=0.5\columnwidth]{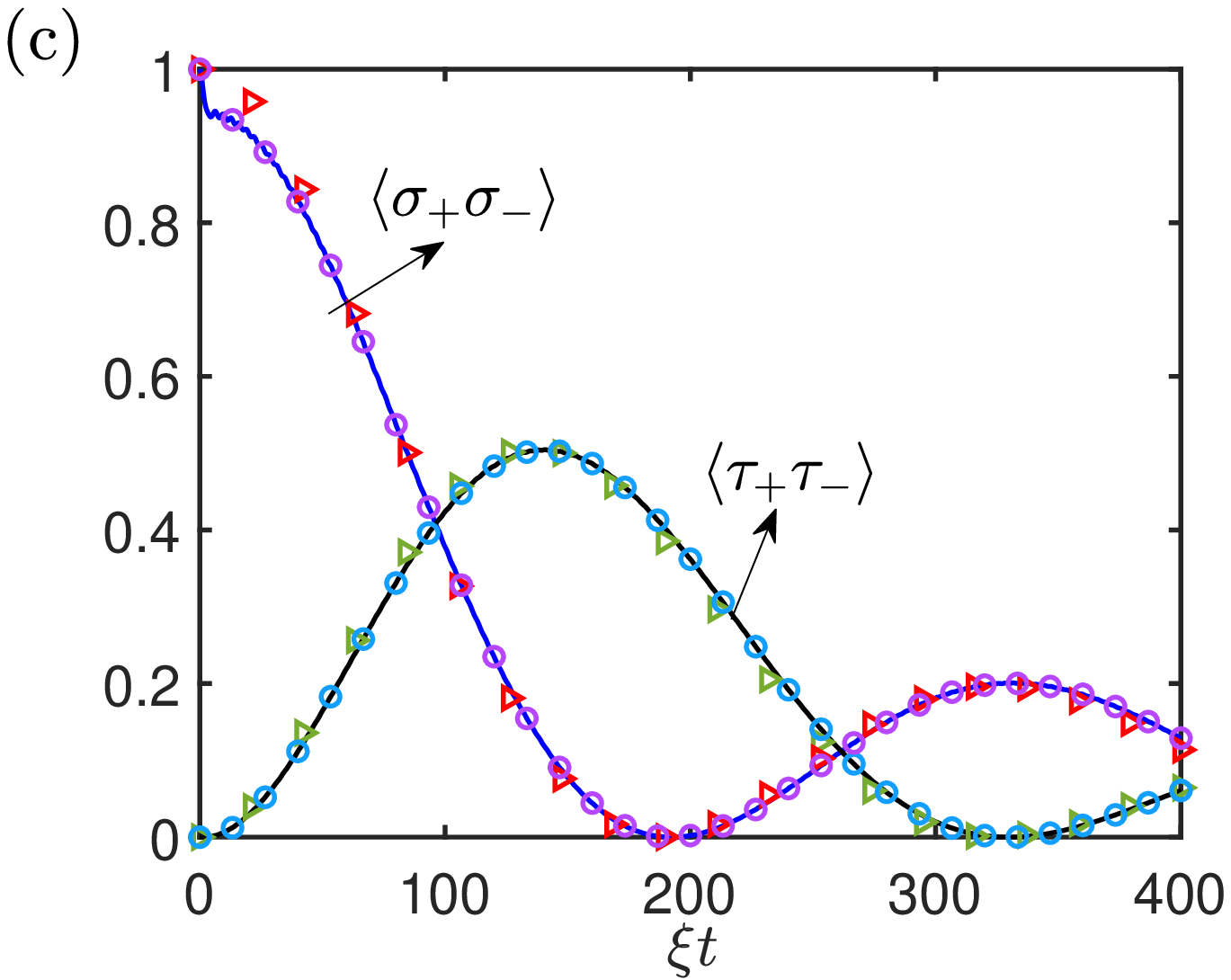}\includegraphics[width=0.5\columnwidth]{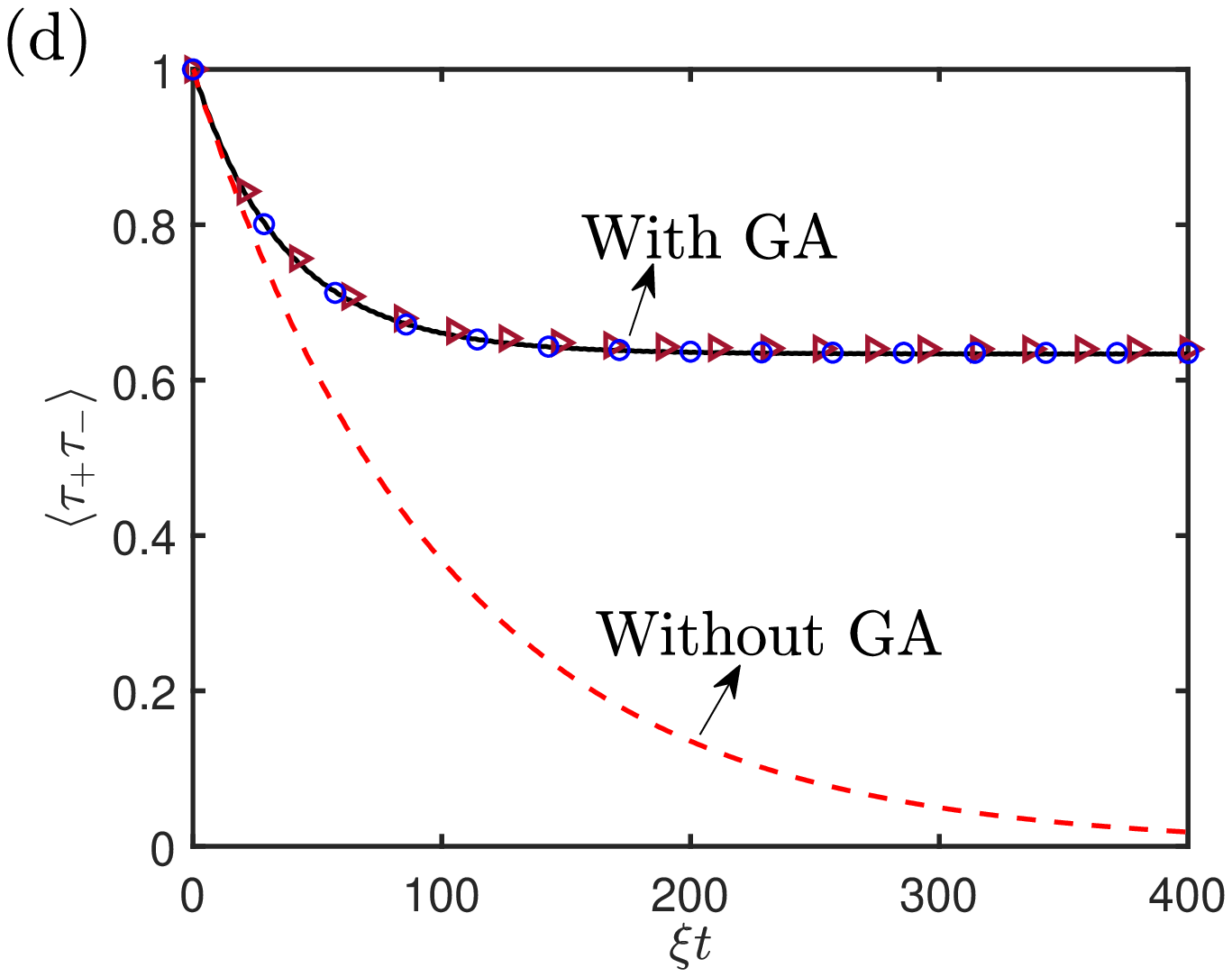}
  \caption{(a) Sketch of the magic cavity QED setup, where an auxiliary small atom is introduced. (b) Excitation of the {giant} atom as a function of the driving detuning $\Delta$.  (c) {Population of the GA $\langle \sigma_+\sigma_-\rangle$ and the small atom $\langle \tau_+\tau_-\rangle$  when the driving field is absent. (d) Population of the small atom with and without the GA. In (c) and (d), the circles represent the numerical result, the triangles are the results of the Markovian approximation master equation Eq.~(12) and the solid lines are obtained based on the non-Markovian approximation Eq.~(19).} The parameters are set as $\Omega=\omega_c=0$, {$\eta=10^{-3}\xi$ in (b)}, $N=6$, $M=1$ in (c), $N=4$, $M=2$ in (d) and $g=g_s=0.1\xi$ in (b-d).}\label{decayx}
\end{figure}

The results within Markovian approximation tell us that the GA with $N=6$ will not dissipate, that is, it forms a perfect magic cavity. In Fig.~4(b), we plot the GA population as a function of the detuning $\Delta$ for different setups. The typical Rabi splitting for $N=6,M=1$ implies that we have achieved the strong coupling $g_I=gg_s/\xi$ in the effective magic cavity QED model. Also, the fact { $\gamma_I=0$} implies that the collective dissipation disappears. So that, we reach a cavity QED model, in which only the small atom undergoes dissipation. We can also observe the Rabi oscillation as shown in Fig.~4(c), where the external driving is dismissed and the initial state is prepared as $|\psi(0)\rangle=\sigma_+|g,g,\rm{vac}\rangle$ in the numerical simulation. It further predicts the valid of the effective magic cavity QED model and the decrease of the oscillation amplitudes are only due to the dissipation of the small atom.

Now, we investigate how the magic cavity QED system behaves for $N=4$, where the GA acts as a leaky magic cavity due to its decay as shown in Fig.~2(a). When the small atom is located in the resonator with $M=1$, it can not effectively couple to the GA ($g_I=\gamma_I=0$). Meanwhile, both of the small atom and GA  emit photon to the CRW and we observe a nearly flat curve for GA population in Fig.~4(b). Alternatively, as for $M=2$, although the small atom and GA can not effectively couple to each other coherently, they undergo collective dissipation to the CRW and a single peak appears when the GA is driven resonantly as shown in Fig.~4(b). Via the collective dissipation, the GA will modulate the dissipation of the small atom. As shown in Fig.~4(d), the small atom undergoes an exponential decay with a characteristic rate $\gamma_s$ in the absence of GA ($g=0$). The role of the GA is clearly demonstrated in Fig.~4(d), {in which we observe a fast initial decay,} which is induced by the collective dissipation between the small atom and GA. After that, the system is trapped in the single excitation subspace without further decay. This rather unexpected feature can be explained by the dark state mechanism. We find that the master equation in the case of $N=4,M=2$ can be simplified as
\begin{eqnarray}
\frac{d\rho}{dt}=2K\rho K^{\dagger} -\rho K^{\dagger}K- K^{\dagger}K\rho
\end{eqnarray}
where $K=\sqrt{\gamma_g}\sigma_{-}+\sqrt{\gamma_s}\tau_{-}$. Therefore, the dark state in the single excitation subspace is expressed as
\begin{equation}
|D\rangle=\frac{(\sqrt{\gamma_s}\sigma_{+}-\sqrt{\gamma_g}\tau_+)
|g,g\rangle}{\sqrt{\gamma_s+\gamma_g}}.
\end{equation}
This explains why the population of the small atom finally achieves the steady value
\begin{equation}
\langle \tau_+\tau_-\rangle(t\rightarrow\infty)=\frac{\gamma_g^2}{(\gamma_s+\gamma_g)^2}.
\end{equation}
for the initial state $|\psi(0)\rangle=\tau_+|g,g\rangle$.
Furthermore, we also numerically find a BIC ($|E^{m}_I\rangle$) in this setup, which is free of decoherence. In terms of this BIC, the final population of the small atom is
\begin{equation}
\langle \tau_+\tau_-\rangle(t\rightarrow\infty)=|\langle \psi(0)|E^{m}_I\rangle\langle E^{m}_I|\tau_+|G\rangle|^2.
\end{equation}
In such a way, we find that the dark state emerges into the BIC, being similar to the two small atom setup, which couples to a common CRW~\cite{An1}. Meanwhile, since $\langle \psi(0)|E^{m}_\alpha \rangle\ll \langle \psi(0)|E^{m}_I\rangle ,\,(\alpha=U,L)$, we find that the BOCs play negligible roles in the dynamics. Therefore, we observe a steady but not oscillation state in the magic cavity setup.

{ For the magic cavity system composed by a two-leg GA and a single small atom as shown in Fig.~4(a), the condition for BIC can be summarized as $KM=m\pi, K(N-M)=n\pi,$ ($m,n\in Z$ and are both odd or both even) where $\Omega=\omega_c-2\xi\cos K$. The case for magic cavity QED model with a multiple-leg giant atom is beyond our consideration in this work.}

Whether the small atom and GA undergo the coherent interaction and collective dissipation can be also explained beyond the Markovian process. To this end, we write the wave function of the magic cavity QED system in the single excitation subspace as
\begin{equation}
  |\psi(t)\rangle=\alpha_g(t)\sigma_{+}|G\rangle+\alpha_s(t)
  \tau_{+}|G\rangle+\sum_k\beta_{k}(t)a_{k}^{\dagger}|G\rangle.
\end{equation}
We set the initial condition as $|\psi(0)\rangle=\tau_+|G\rangle$, and the dynamical equations for $\alpha_g$ and $\alpha_s$ can be obtained as (see Appendix C)
\begin{eqnarray}
\dot{\alpha}_g(t)&=&M_{gg}(t)\alpha_g(t)+M_{gs}(t)\alpha_s(t),\nonumber \\
\dot{\alpha}_s(t)&=&M_{gs}(t)\alpha_g(t)+M_{ss}(t)\alpha_s(t),
\end{eqnarray}
where
\begin{eqnarray}
M_{gg}&=&-2g^2\int_0^{t}d\tau G(\tau),M_{ss}=-g_s^2\int_0^td\tau J_0(2\xi\tau),\nonumber \\M_{gs}&=&-gg_s\int_{0}^{t}d\tau Q(\tau).
\end{eqnarray}
Therefore, the information of the interaction between the small atom and GA can be extracted from
\begin{eqnarray}
Q(\tau)=i^MJ_M(2\xi\tau)+i^{N-M}J_{N-M}(2\xi\tau).
\end{eqnarray}
For the case of $N=6, M=1$, we will reach $Q(\tau)=i[J_1(2\xi\tau)+J_5(2\xi\tau)]$, in which the constructive interference of the terms leads to a strong interaction, and we can observe a Rabi splitting and oscillation. Similarly, for the case of $N=4$, the constructive interference for $M=2$ and destructive interference for $M=1$ leads to the dramatically different results, which are shown in Fig.~4(b).

{We note that, in Fig.~4(c,d), we also give the comparison of the results of numerical, Markovian approximation and non-Markovian approximation. The three results agree with each other in the considered parameter regime. It means that the interference effect as well as the BIC plays key role in the magic cavity QED system.}

\section{Discussion and conclusions}

In summary, we have proposed a magic cavity realized by the on demand GA, which couples to a CRW via two connecting points. Such GA traps the emitted photons between the coupling points via BIC-BOC interference mechanism. We further proposed an effective magic cavity QED setup, which can be tuned from the perfect cavity to leaky cavity, and therefore overcome the difficulty of non-adjustability in the real cavity QED scenarios. In the microwave domain, the GA has been realized by coupling the transmon qubit to the transmission line. In such systems, the parameters can be achieved in the regime $g,g_s\leq\xi\approx50-200$\,MHz with the existing technology~\cite{xi1,xi2,xi3}. Alternatively, the single small atom can also be implemented by the superconducting qubits~\cite{sa}. The single small atom can also be replaced by an ensemble of Rydberg atoms to enhance the light-matter interaction and demonstrate the effects in magic cavity QED model which is predicted in this paper.

In the previous studies, it is shown that the bound state in the open system is helpful for preventing decoherence and beneficial for quantum precise measurement~\cite{An2,An3}. We here further exhibit how the interference effect among different kinds of bound states modify the dynamics of a quantum system in the structured environment, and can be developed to more complex waveguide setup, or to investigate the many body physics.

\begin{acknowledgments}
We thank the fruitful discussion with Prof. Peter Rabl.
Z. W. is supported by National Key R\&D Program of China (Grant
No. 2021YFE0193500) and Science and Technology Development Project of Jilin Province (Grant No. 20230101357JC). Z. G. is supported by National Science foundation of China (Grant No.12175150) and the nature science foundtion of Guang-dong Province (Grant Nos. 2019A1515011400 and 2023A1515011223).
\end{acknowledgments}

\appendix%\appendixpage
\addcontentsline{toc}{section}{Appendices}\markboth{APPENDICES}{}
\begin{subappendices}
\begin{widetext}
\section{Bound state in the continuum (BIC)}

In this appendix, we discuss the property and existence condition of the BIC.

We first consider that a single giant atom couples to the CRW via the $0$th and $N$th resonator, the Hamiltonian is written as ($\hbar=1$)
\begin{equation}
H_{\rm GA}=\omega_c \sum_{j}a_{j}^{\dagger}a_{j}-\xi \sum_{j}(a_{j+1}^{\dagger}a_{j}+a_{j}^{\dagger}a_{j+1})+\Omega \sigma_+\sigma_-+g[(a_{0}^{\dagger}+a_{N}^{\dagger})\sigma_- +(a_{0}+a_{N})\sigma_+ ].
\end{equation}
Working in the momentum space, we introduce the Fourier transformation  $a_j=\sum_k a_k e^{ikj}/\sqrt{N_c}$, with $N_c=\infty$ being the length of the CRW, then Hamiltonian becomes $H_{GA}=H_0+H_I$, where
\begin{eqnarray}
H_0&=&\sum_k\omega_k a_k^\dagger a_k+\Omega|e\rangle\langle e|,\label{kspace1}\\
H_I&=&\frac{g}{\sqrt{N_c}}\sum_k
[(1+e^{ikN})a_k^{\dagger}\sigma_-+{\rm H.c.}],
\label{kspace2}
\end{eqnarray}
where $\omega_k=\omega_c-2\xi\cos k$ is the dispersion relation of the CRW.

Based on the Hamiltonian in Eqs.~(\ref{kspace1},\ref{kspace2}), we can obtain the bound state of the atom-waveguide coupled system. To this end, we assume the eigen state $|E\rangle$ with eigen energy $E$ in the single excitation subspace as
\begin{equation}
|E\rangle=(\sum_k b_k {a_{k}^\dagger}+ c \sigma_+)|g,{\rm vac}\rangle,
\end{equation}
then, the Sch\"{o}dinger equation $H|E\rangle=E|E\rangle$ yields the coupled equation
\begin{eqnarray}
cE&=&\frac{g}{\sqrt{N_c}}\sum_k b_k(1+e^{-ikN}),\\(E+2\xi\cos k)b_k&=&\frac{g}{\sqrt{N_c}} c(1+e^{ikN}),\label{bk}
\end{eqnarray}
where we have set $\omega_c=\Omega=0$. As a result, eliminating the photonic amplitudes $b_k$, we will obtain the transcendental equation for the eigen energy as
\begin{equation}
E=\frac{g^2}{\pi}\int_{-\pi}^{\pi} dk\frac{1+\cos kN}{E+2\xi\cos k}.
\end{equation}
Ref.~\cite{BOC2} shows that one can always obtain two BOCs, in which the photon is mainly populated on the atom-waveguide coupling sites. We also find the interesting BIC with $E=0$ for $N=4m+2,\,m\in Z$. In this case, we can obtain from Eq.~(\ref{bk}) as
\begin{equation}
\frac{b_k}{c}=\frac{g(1+e^{ikN})}{2\xi \sqrt{N_c}\cos k}.
\end{equation}
Furthermore, we can extract the photonic distribution in the real space as
\begin{equation}
\frac{b_j}{c}=\frac{1}{\sqrt{N_c}}\sum_k\frac{b_k}{c}e^{-ikj}=
\sum_k\frac{g(1+e^{ikN})e^{ikj}}{2\xi N_c\cos k}
={\frac{g}{4\pi\xi}}\int_{-\pi}^{\pi}dk\frac{(1+e^{ikN})e^{ikj}}{\cos k}.
\end{equation}
\begin{figure}
  \centering
  \includegraphics[width=0.33\columnwidth]{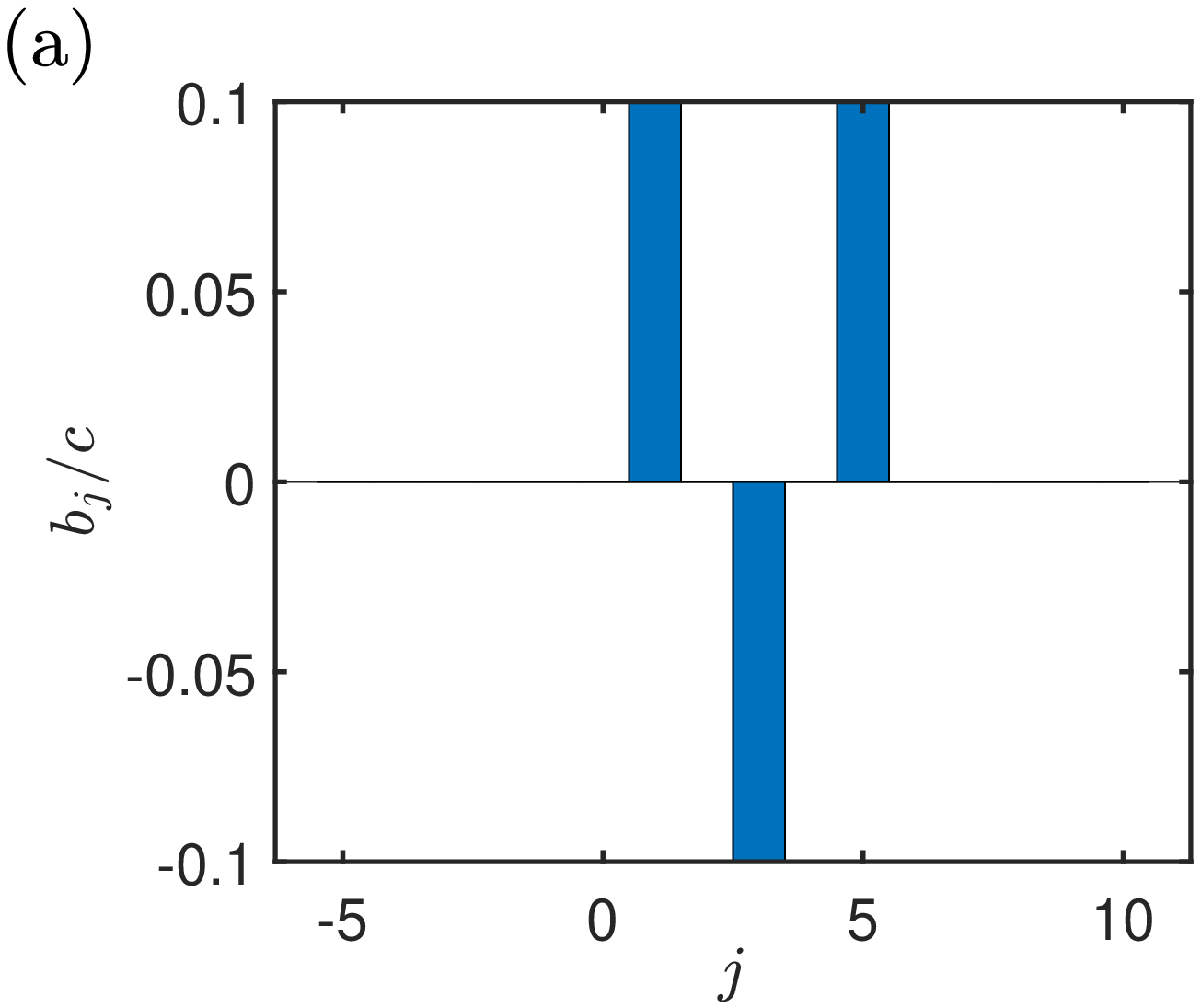}\includegraphics[width=0.33\columnwidth]{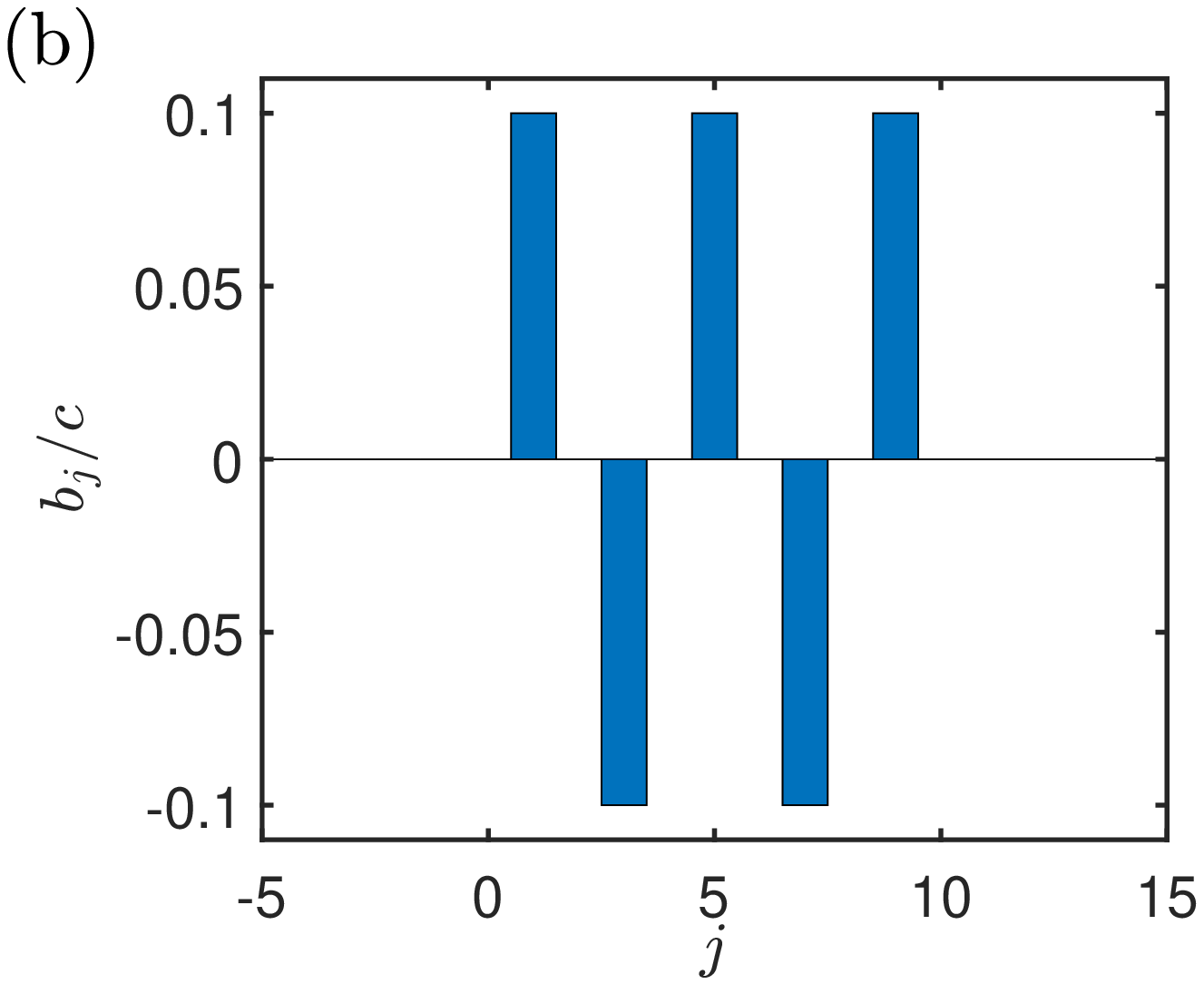}\includegraphics[width=0.33\columnwidth]{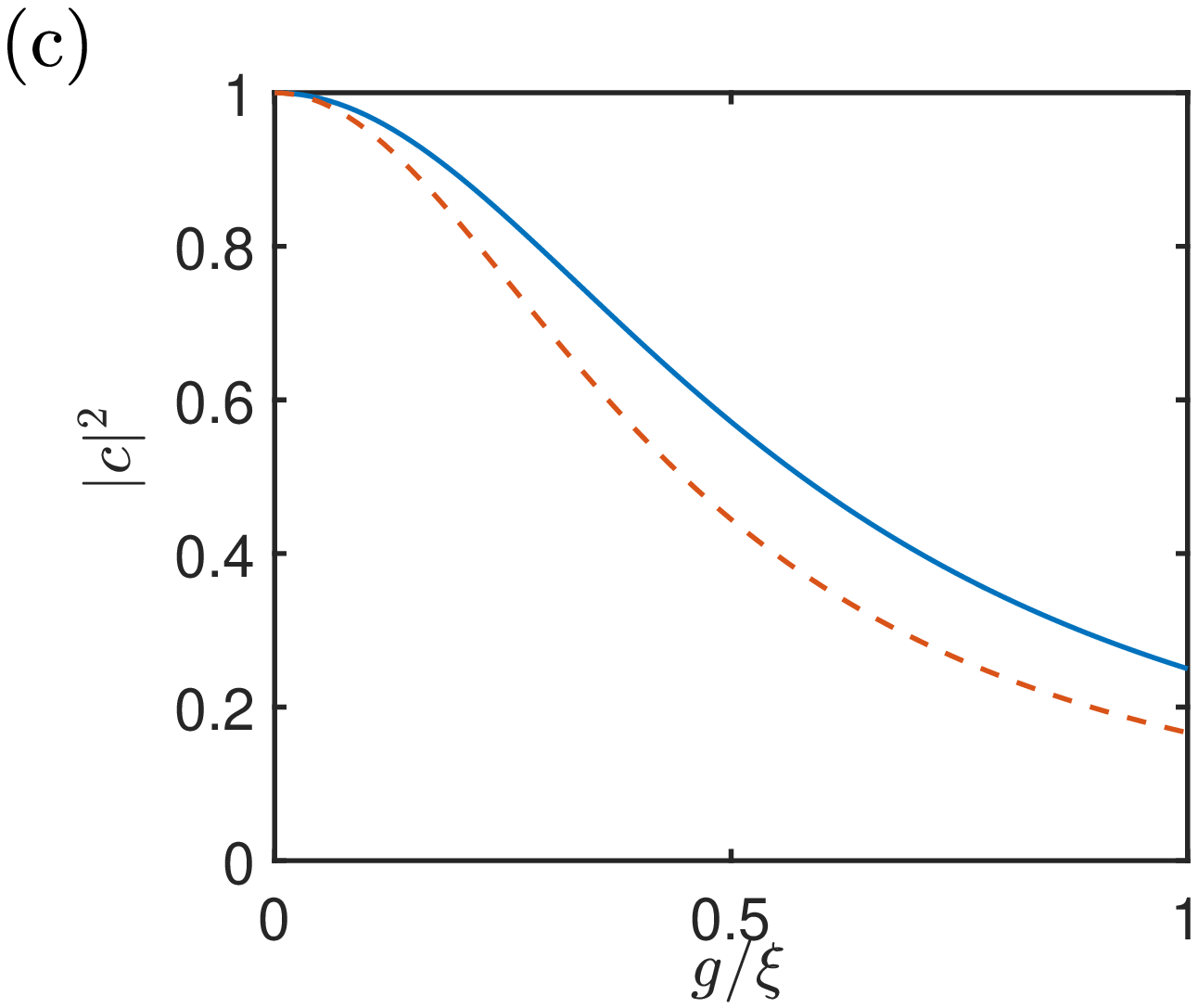}
  \caption{Photonic distribution (a, b) and atomic excitation (c) in the BIC.
   The parameters are set as $\Omega=\omega_c=0$ and $N=6,g=0.1\xi$ for (a) and $N=10,g=0.1\xi$ for (b). In Fig.~(c), the solid (dashed) line represents $N=6\,(10)$. }
   \label{ss1}
\end{figure}
\begin{figure}
  \centering
  \includegraphics[width=0.33\columnwidth]{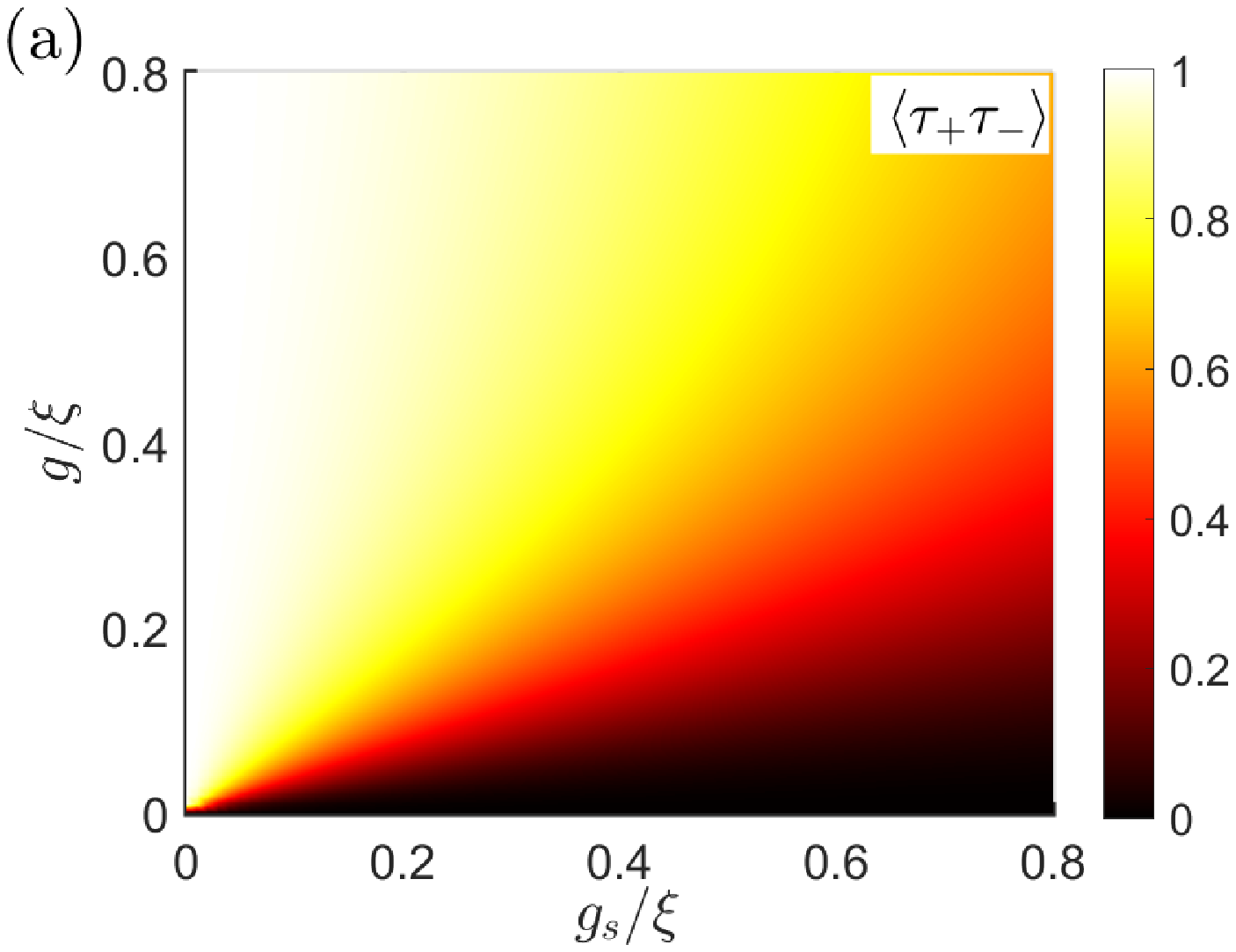}\includegraphics[width=0.33\columnwidth]{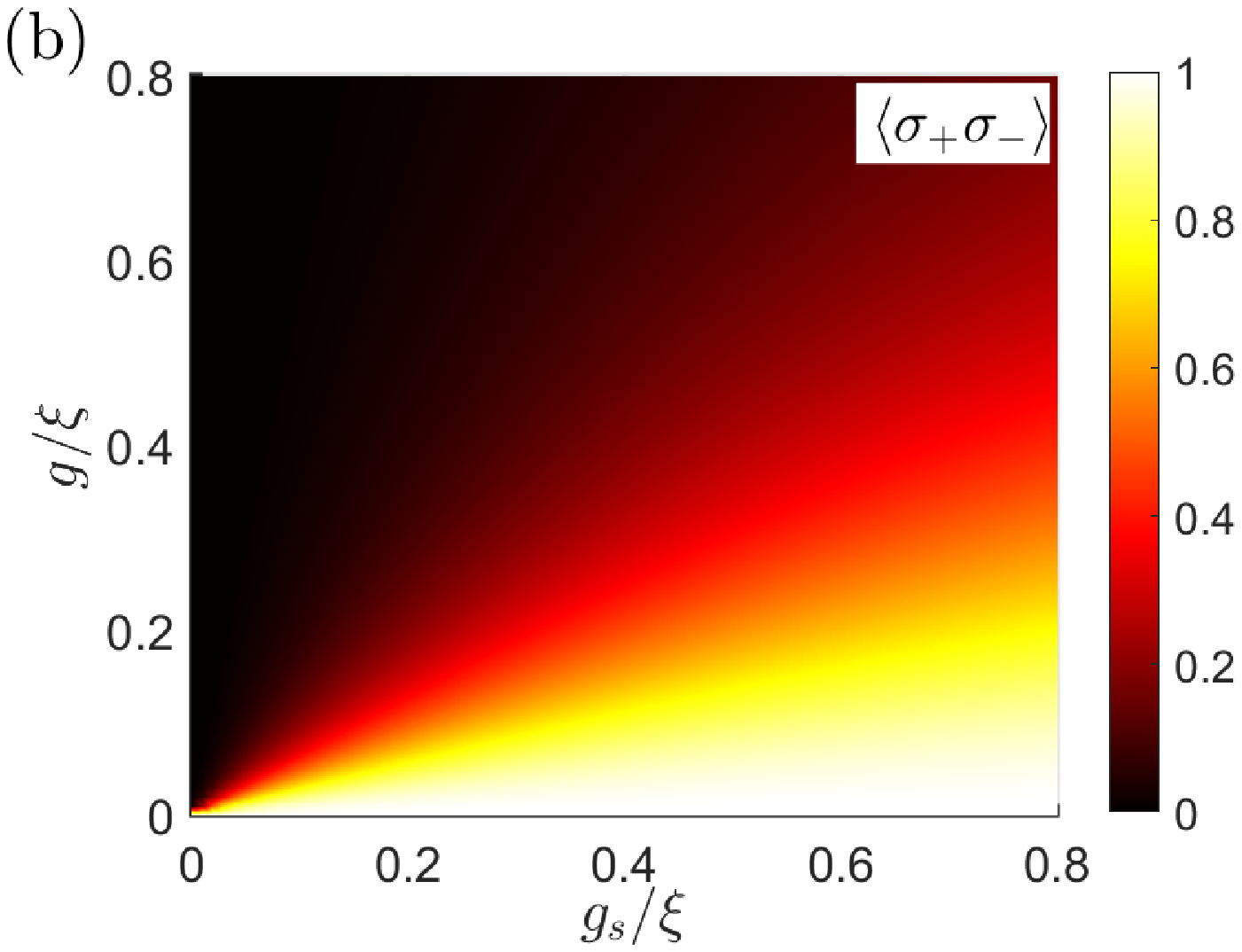}\includegraphics[width=0.33\columnwidth]{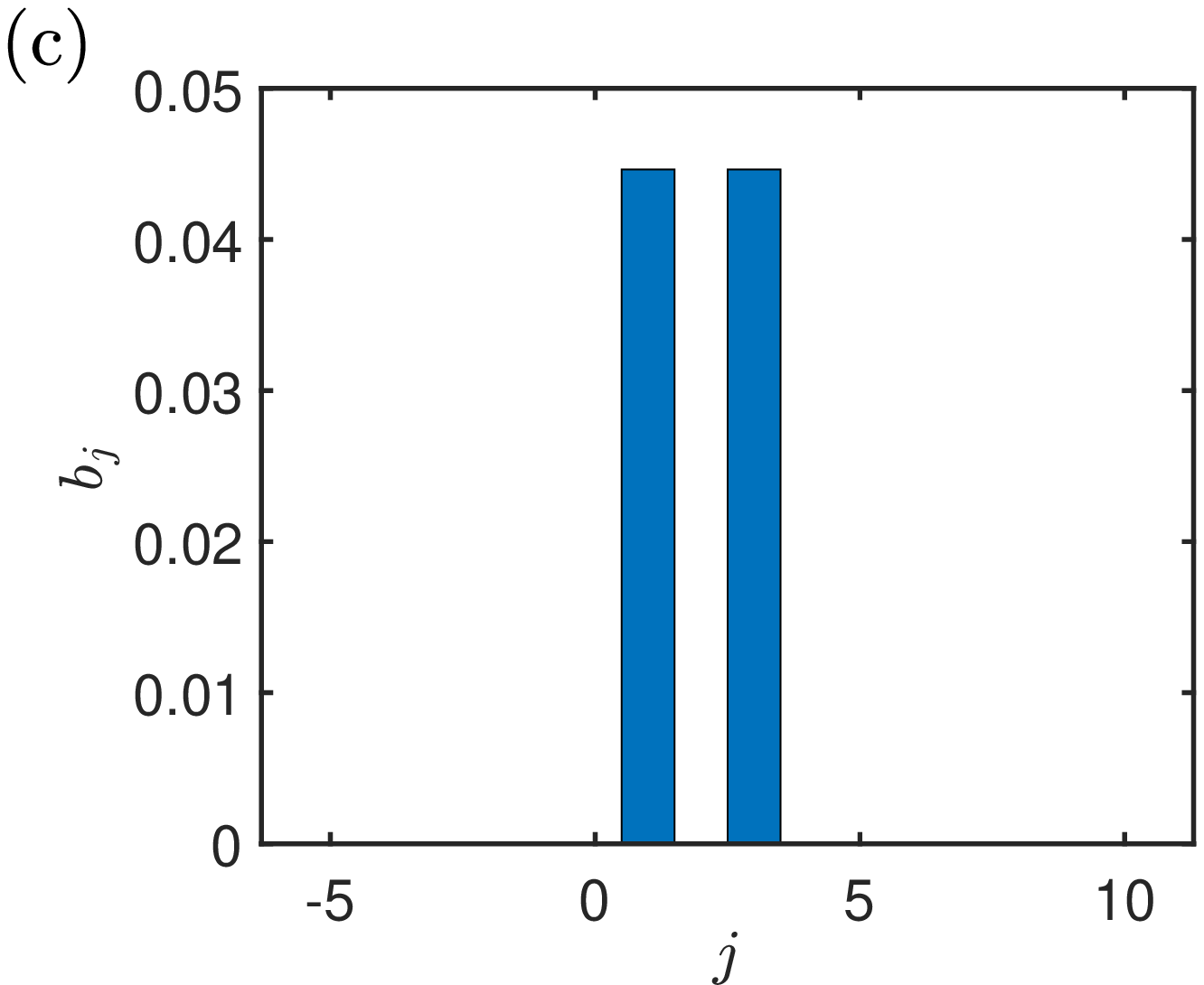}
  \caption{ Population of the small atom (a) and giant atom (b) for the BIC in the magic cavity QED setup.  (c) Photon distribution for the BIC when $g=g_s=0.1\xi$. The parameters are set as $N=4,M=2$ and $\Omega=\omega_c=0$.}
  \label{ss2}
\end{figure}
In Fig.~\ref{ss1} (a) and (b), we show the histogram for the photonic distribute for $N=6$ and $N=10$, respectively. It shows that the photon is bounded inside the giant atom ($b_j=0$ for $j>N$ or $j<0$). Moreover, the photons only populate uniformly the sites with odd number for $j=1,3,5\cdots$. Meanwhile, we find that the value of $b_j/c$ increases with the coupling strength $g$, therefore, the atomic population decreases after normalization as shown in ~\ref{ss1}(c).

For the magic QED system, we also find a BIC for $N=4,M=2$, in which the population for small atom $\langle \tau_+\tau_-\rangle$ and giant atom $\langle \sigma_+\sigma_-\rangle$ are plotted as a function of coupling strength $g$ and $g_s$ in Fig.~\ref{ss2}(a, b). {For $g=0.1\xi,g_s=0.1\xi$, we plot the photonic population for the BIC in Fig.~\ref{ss2}(c), which shows that the photons are only bound in the first and third sites}.

The above discussions and the results in the main text show that the BIC exist in the single two-leg giant atom system when $\Omega=\omega_c, N=6$ and  magic QED system when $\Omega=\omega_c, N=4, M=2$. {We also find that the BIC is always present when $\sum_{1}^{M}\exp(iKn_j)=0$ in a single giant atom with more than two legs, where $K$ satisfies $\Omega=\omega_c-2\xi\cos K$ and $n_j$ is the position of the $jth$ coupling point in the coupled resonator waveguide and $M$ is the number of the total coupling points between the giant atom and the CRW, this result agrees with that given in Ref.~\cite{limpra} . Moreover, for the magic cavity system composed by a two-leg giant atom and a single small atom as shown in Fig.~4(a) of the main text, the condition for BIC can be summarized as $KM=m\pi, K(N-M)=n\pi,$ ($m,n\in Z$ and are both odd or both even) where $\Omega=\omega_c-2\xi\cos K$. The case for magic cavity QED model with a multiple-leg giant atom is beyond our consideration in this work.}

\section{Master equation under Markovian approximation}
In this section of the supplementary material (SM), we derive the master equation which governs the dynamics of the system by considering the coupled resonator waveguide (CRW) as the structured environment, the similar calculation can also be found in Ref.~\cite{wei}.

In the interaction picture, the interaction Hamiltonian is written as
 \begin{equation}
H_I(t)=g\sum_{i=1}^{2}[\sigma^{+}E(n_{i},t)e^{i\Omega t}+\sigma^{-}E^{\dagger}(n_{i},t)e^{-i\Omega t}]
\label{interaction}
\end{equation}
where $E(n_{i},t)=\frac{1}{\sqrt{N_{c}}}\sum_{k}(e^{-i\omega_{k}t}e^{ikn_{i}}a_{k})$ and $n_1=0,n_2=N$. Under the Markov approximation and working in the interaction picture, the formal master equation for a quantum open system reads
{\begin{eqnarray}
\dot{\rho}(t)&=&-\int_{0}^{\infty}d\tau{\rm Tr}_{c}\{[H_{I}(t),[H_{I}(t-\tau),\rho_{c}\otimes\rho(t)]]\}\nonumber \\
\end{eqnarray}}	
Since we are working at the zero temperature, the CRW is in the vacuum state initially, therefore, we will have ${\rm Tr}_{c}[E^{\dagger}(n_{i},t)E(n_{j},t-\tau)\rho_{c}]=0$, and the above equation becomes (going back to the Schr\"{o}dinger picture)
\begin{equation}
\dot{\rho}=-i\Omega[|e\rangle\langle e|,\rho]+(A+A^{*})\sigma^{-}\rho\sigma^{+}
-A\sigma^{+}\sigma^{-}\rho-A^{*}\rho\sigma^{+}\sigma^{-}
\end{equation}
where~\cite{Peter16pra}
\begin{eqnarray}
A&=&g^2\int_{0}^{\infty}d\tau e^{i\Omega\tau}{\rm Tr}_{c}[\sum_{i,j}E(n_{i},t)E^{+}(n_{j},t-\tau)\rho_{c}]\nonumber\\
	&=&	g^2\sum_{i,j}\int_{0}^{\infty}d\tau e^{i\Omega\tau}{\rm Tr}[E(n_{i},t)E^{+}(n_{j},t-\tau)\rho_{c}]\nonumber \\
	&=&	g^2\sum_{i,j}\int_{0}^{\infty}d\tau \frac{e^{i\Omega\tau}}{N_{c}}\times{\rm Tr}[\sum_{k,k'}e^{-i\omega_{k}t}e^{ikn_{i}}a_{k}e^{i\omega_{k'}(t-\tau)}e^{-ik'n_{j}}
a_{k'}^{\dagger}\rho_{c}]\nonumber \\
	&=&	g^2\sum_{i,j}\int_{0}^{\infty}d\tau\frac{1}{N_{c}}\sum_{k}[e^{-i(\omega_{k}-\Omega)\tau}
e^{-ik(n_{j}-n_{i})}]\nonumber \\
	&=&	g^2\sum_{i,j}\int_{0}^{\infty}d\tau\frac{1}{N_{c}}\sum_{n=0}^{N_{c}-1}e^{-i\Delta_c\tau}e^{\frac{-2\pi i(n_{j}-n_{i})n}{N_{c}}}e^{2i\xi\cos(\frac{2\pi n}{N_{c}})\tau}\nonumber\\
	&=&	g^2\sum_{i,j}\int_{0}^{\infty}d\tau\frac{e^{-i\Delta_c\tau}}{N_{c}}\sum_{n=0}^{N_{c}-1}e^{\frac{-2\pi i(n_{j}-n_{i})n}{N_{c}}}\times\sum_{m=-\infty}^{\infty}i^{m}J_{m}(2\xi\tau)e^{i2\pi nm/N_{c}}\nonumber \\
	&=&	g^2\sum_{i,j}\int_{0}^{\infty}d\tau e^{-i\Delta_c\tau}i^{|n_{i}-n_{j}|}J_{|n_{i}-n_{j}|}(2\xi\tau)\nonumber \\
	&=&	g^2\sum_{i,j}\frac{1}{2\xi}e^{\frac{i\pi|n_{i}-n_{j}|}{2}}\nonumber \\&=&\frac{g^2}{\xi}(1+e^{i\pi|n_{1}-n_{2}|/2}).
\end{eqnarray}
In the above calculations, we have considered that the giant atom is resonant with the bare cavity ($\Delta_c:=\omega_c-\Omega=0$), and used the formula
\begin{equation}
\int_{0}^{\infty}d\tau J_{m}(a\tau)=\frac{1}{|a|}.
\end{equation}

Back to the configuration we are considering ($n_1=0,n_2=N$), we finally obtain \begin{equation}
A=\frac{g^2}{\xi}(1+e^{i\pi N/2}).
\end{equation}
Therefore, we will have $A=2g^2/\xi$ for $N=4$, in which the giant atom undergoes an exponential decay. More interestingly, when $N=6$, we will have $A=0$, which implies the giant atom will not decay within the Markovian approximation.

Then, we further consider the magic cavity QED system where an additional small atom couples to the $M$th ($0<M<N$) resonator in CRW, the interaction Hamiltonian in the momentum space is written as
 \begin{equation}
H_I(t)=g\sum_{i=1}^{2}[\sigma_{+}E(n_{i},t)e^{i\Omega t}+\sigma_{-}E^{\dagger}(n_{i},t)e^{-i\Omega t}]+g_s[\tau_{+}E(m,t)e^{i\Omega t}+\tau_{-}E^{\dagger}(m,t)e^{-i\Omega t}].
\label{interactionc}
\end{equation}
Following the similar derivations as those for a single giant atom, the master equation for the magic cavity QED model can be obtained as
\begin{eqnarray}
\frac{d\rho}{dt} & = &-i\Omega[\sigma_+\sigma_-+\tau_+\tau_-,\rho]\nonumber \\&&+
(A_1+A_1^{*})\sigma_{-}\rho\sigma_{+}-
A_{1}\sigma_{+}\sigma_{-}\rho-A_{1}^{*}\rho\sigma_{+}\sigma_{-}+
A_{2}[2\tau_{-}\rho\tau_{+}-\tau_{+}\tau_{-}\rho-\rho\tau_{+}\tau_{-}]\nonumber \\
 &&+(B+B^{*})(\sigma_{-}\rho\tau_{+}+\tau_{-}\rho\sigma_{+})
 -B(\tau_{+}\sigma_{-}\rho+\tau_{-}\sigma_{+}\rho)
 -B^{*}(\rho\tau_{+}\sigma_{-}+\rho\tau_{-}\sigma_{+}).
\end{eqnarray}
where
\begin{eqnarray}
A_{1}=\frac{g^2(1+i^N)}{\xi},\,A_2=\frac{g_s^2}{2\xi},\,
B=\frac{gg_s}{2\xi}[i^M+i^{(N-M)}].
\end{eqnarray}

When the external driving is taken into consideration phenomenologically, we should work in the rotating frame to eliminate the time dependent in the Hamiltonian. After regrouping some individual terms, we will get the final form, which is given in {Eq.~(\ref{master})} of the main text.

\section{Non-Markovian dynamics}

In this section, we give the non-Markovian amplitude equations for both of the single giant atom and magic cavity QED system.

For the single giant atom, we assume the wave function at time $t$ is given by
 \begin{equation}
|\psi(t)\rangle=e^{-i\Omega t}\left(\alpha (t)\sigma_++\sum_{k}\beta_{k}(t)a_{k}^{\dagger}\right)|g,{\rm vac}\rangle.
\end{equation}
 Governed by the Hamiltonian in Eqs.~(\ref{kspace1},\ref{kspace2}), we will have
 \begin{eqnarray}
i\frac{\partial}{\partial t}\alpha & = & \frac{g}{\sqrt{N_{c}}}\sum_{k}\beta_{k}(1+e^{-iNk}),\label{eq:alpha}\\
i\frac{\partial}{\partial t}\beta_{k} & = & \Delta_{k}\beta_{k}+\frac{g}{\sqrt{N_{c}}}(1+e^{ikN})\alpha,\label{eq:beta}
\end{eqnarray}
where $\Delta_{k}=\omega_{k}-\Omega.$ Then, in the condition of $\beta_{k}(0)=0,\alpha(0)=1$, we will have
\begin{equation}
\beta_{k}=-\frac{ig}{\sqrt{N_{c}}}(1+e^{ikN})\int_{0}^{t}d\tau\alpha(\tau)e^{-i\Delta_{k}(t-\tau)}.\label{eq:betak}
\end{equation}
As a result, Eq.~(\ref{eq:alpha}) becomes
\begin{eqnarray}
\frac{\partial}{\partial t}\alpha & = & -\frac{g^{2}}{N_{c}}\sum_{k}\left[(1+e^{ikN})(1+e^{-ikN})\int_{0}^{t}d\tau\alpha(\tau)e^{-i\Delta_{k}(t-\tau)}\right]\nonumber \\
 & = & -\frac{g^{2}}{2\pi}\int_{-\pi}^{\pi}dk\int_{0}^{t}d\tau(1+e^{ikN})(1+e^{-ikN})\alpha(\tau)e^{-i\Delta_{k}(t-\tau)}\nonumber \\
 & = & -\frac{g^{2}}{2\pi}\int_{-\pi}^{\pi}dk\int_{0}^{t}d\tau(1+e^{ikN})(1+e^{-ikN})\alpha(\tau)e^{2i\xi\cos k(t-\tau)}\nonumber \\
 & = & -\frac{g^{2}}{2\pi}\int_{-\pi}^{\pi}dk\int_{0}^{t}d\tau(2+e^{ikN}+e^{-ikN})\alpha(\tau)e^{2i\xi\cos k(t-\tau)}\nonumber \\
 & = & -g^{2}\int_{0}^{t}d\tau\alpha(\tau)\left[\frac{1}{\pi}\int_{-\pi}^{\pi}dke^{2i\xi\cos k(t-\tau)}+\frac{1}{2\pi}\int_{-\pi}^{\pi}dke^{i[2\xi\cos k(t-\tau)+kN]}+\frac{1}{2\pi}\int_{-\pi}^{\pi}dke^{i[2\xi\cos k(t-\tau)-kN]}\right].\nonumber \\
\end{eqnarray}
By use of the formula
\begin{equation}
e^{iz\cos\theta}=\sum_{n=-\infty}^{n=\infty}i^{n}J_{n}(z)e^{in\theta},\,
\int_{-\pi}^{\pi}e^{i(n-m)k}dk=2\pi\delta_{n,m},\,J_{-N}(x)=(-1)^{N}J_{N}(x),
\end{equation}
we will have
\begin{equation}
\frac{\partial}{\partial t}\alpha(t)=-2g^{2}\int_{0}^{t}d\tau\alpha(\tau)\left\{ J_{0}[2\xi(t-\tau)]+i^{N}J_{N}[2\xi(t-\tau)]\right\} \label{eq:alphat}
\end{equation}
We now further perform the Weisskopf-Wigner approximation to replace $\alpha(\tau)$ by $\alpha(t)$ in Eq.(\ref{eq:alphat}), then we will have
\begin{eqnarray}
\frac{\partial}{\partial t}\alpha(t) &
\approx & -2g^{2}\alpha(t)\int_{0}^{t}d\tau \left\{ J_{0}[2\xi(t-\tau)]+i^{N}J_{N}[2\xi(t-\tau)]\right\} \nonumber \\
 & = & -2g^{2}\alpha(t)\int_{0}^{t}d\tau \left[ J_{0}(2\xi\tau)+i^{N}J_{N}(2\xi\tau)\right] .
\end{eqnarray}
Therefore, the solution of $\alpha(t)$ can be obtained as
\begin{equation}
\alpha(t)=e^{-2g^2\int_{0}^{t}dt_1 \int_{0}^{t_1}d\tau [J_{0}(2\xi\tau)+i^{N}J_{N}(2\xi\tau)]},
\end{equation}
which is {Eq.~(\ref{exp})} in the main text.
To find the dynamics of the photon distribution, we should perform
the inverse Fourier transformation. Since
\begin{equation}
\sum_{k}\beta_{k}a_{k}^{\dagger}|0g\rangle=\frac{1}{\sqrt{N_{c}}}\sum_{k,j}\beta_{k}a_{j}^{\dagger}e^{-ikj}|0g\rangle\equiv\sum_{j}\beta_{j}a_{j}^{\dagger}|0g\rangle,
\end{equation}
we have
\begin{equation}
\beta_{j}=\frac{1}{\sqrt{N_{c}}}\sum_{k}\beta_{k}e^{-ikj}.
\end{equation}
Then, combining Eq.(\ref{eq:betak}), we will have
\begin{eqnarray}
\beta_{j} & = & -\frac{ig}{N_{c}}\sum_{k}(1+e^{ikN})\int_{0}^{t}d\tau\alpha(\tau)
e^{-i\Delta_{k}(t-\tau)}e^{-ikj}\nonumber\\
 & = & -\frac{ig}{2\pi}\int dk(1+e^{ikN})\int_{0}^{t}d\tau\alpha(\tau)e^{-i\Delta_{k}(t-\tau)}e^{-ikj}\nonumber \\
 &=& -\frac{ig}{2\pi}\int_{0}^{t}d\tau\alpha(\tau){\mathcal F}_j(t-\tau)
 \nonumber \\
 &=& -\frac{ig}{2\pi}\int_{0}^{t}d\tau\alpha(t-\tau){\mathcal F}_j(\tau)
\end{eqnarray}
where
\begin{eqnarray}
\mathcal{F}_j(\tau) & = & \int dke^{-i\Delta_{k}\tau}(1+e^{ikN})e^{-ikj}\nonumber \\
 & = &\int dke^{i\tau2\xi\cos k}(1+e^{ikN})e^{-ikj}\nonumber \\
 & = &\int dk\sum_{n}i^{n}J_{n}(2\xi\tau)e^{ink}(e^{-ikj}+e^{ik(N-j)})\nonumber \\
 & = & 2\pi \left[ i^{j}J_{j}(2\xi\tau)+i^{(j-N)}J_{j-N}(2\xi\tau)\right] ,
\end{eqnarray}
Therefore, we will have
\begin{equation}
\beta_{j}(t)=-ig\int_{0}^{t}d\tau\alpha(t-\tau)\left[ i^{j}J_{j}(2\xi\tau)+i^{(j-N)}J_{j-N}(2\xi\tau)\right],
\end{equation}
which is {Eq.~(\ref{beta})} in the main text.

As for the magic cavity QED system, the Hamiltonian is given in Eq.~(\ref{interactionc}), following the same process as above, we will obtain the amplitude equation for the wave function
\begin{equation}
  |\psi(t)\rangle=\alpha_g(t)\sigma_{+}|G\rangle+\alpha_s(t)\tau_{+}|G\rangle+\sum_k\beta_{k}(t)a_{k}^{\dagger}|G\rangle
\end{equation}
as
\begin{eqnarray}
\dot{\alpha}_g=&&-gg_s\alpha_{s}(t)\int_{0}^{t}d\tau
\left[i^MJ_M(2\xi\tau)+i^{N-M}J_{N-M}(2\xi\tau)\right]
-2g^{2}\alpha_g(t)\int_{0}^{t}d\tau
\left[J_0(2\xi\tau)+i^{N}J_{N}(2\xi\tau)\right],\\
\dot{\alpha}_s=&&-gg_s\alpha_{g}(t)\int_{0}^{t}d\tau
\left[i^MJ_M(2\xi\tau)+i^{N-M}J_{N-M}(2\xi\tau)\right]-g_{s}^{2}\alpha_s(t)\int_{0}^{t}d\tau J_0(2\xi\tau)
\label{magic}
\end{eqnarray}
which are actually {Eqs.~(19-21)} in the main text.
\end{widetext}
\end{subappendices}


\begin{thebibliography}{99}
\bibitem{S1}M. V. Gustafsson, T. Aref, A. F. Kockum, M. K. Ekstr\"om, G. Johansson, and P. Delsing, Propagating phonons coupled to an artificial atom, Science \textbf{346}, 207 (2014).

\bibitem{S2}L. Guo, A. Grimsmo, A. F. Kockum, M. Pletyukhov, and G. Johansson, Giant acoustic atom: A single quantum system with a deterministic time delay, Phys. Rev. A \textbf{95}, 053821 (2017).

\bibitem{sq1}G. Andersson, B. Suri, L. Guo, T. Aref, and P. Delsing, Non-exponential decay of a giant artificial atom, Nat. Phys. \textbf{15}, 1123 (2019).

\bibitem{sq2}B. Kannan, M. J. Ruckriegel, D. L. Campbell, A. F. Kockum, J. Braum\"uller, D. K. Kim, M. Kjaergaard, P. Krantz, A. Melville, B. M. Niedzielski, A. Veps\"al\"ainen, R. Winik, J. L. Yoder, F. Nori, T. P. Orlando, S. Gustavsson, and W. D. Oliver, Waveguide quantum electrodynamics with superconducting artificial giant atoms, Nature \textbf{583}, 775 (2020).

\bibitem{sq3}A. M. Vadiraj, A. Ask, T. G. McConkey, I. Nsanzineza, C. W. Sandbo Chang, A. F. Kockum, and C. M. Wilson, Engineering the level structure of a giant artificial atom in waveguide quantum electrodynamics, Phys. Rev. A, {\bf 103}, 023710 (2021).

\bibitem{mq}Z. Q. Wang, Y. P. Wang, J. Yao, R. Shen, W. Wu, J. Qian, J. Li, S. Zhu, and J. Q. You, Giant spin ensembles in waveguide magnonics, Nat. Commun. {\bf 13}, 7580 (2022).

\bibitem{g5}A. F. Kockum, Quantum optics with giant atoms-the first five years, in International Symposium on Mathematics, Quantum Theory, and Cryptography, (Springer Singapore, Singapore, 2021), p. 125.

\bibitem{Lambpra14}A. F. Kockum, P. Delsing, and G. Johansson, Designing frequency-dependent relaxation rates and Lamb shifts for a giant artificial atom, Phys. Rev. A \textbf{90}, 013837 (2014).

\bibitem{ar2022}S. Terradas-Brians\'o, C. A. Gonz\'alez-Guti\'errez, F. Nori, L. Mart\'{\i}n-Moreno, and D. Zueco, Ultrastrong waveguide QED with giant atoms, Phys. Rev. A {\bf 106}, 063717 (2022).

\bibitem{AF2018}A. F. Kockum, G. Johansson and F. Nori, Decoherence-free interaction between giant atoms in waveguide quantum electrodynamics, Phys. Rev. Lett.  {\bf 120}, 140404 (2018).

\bibitem{AC2020}A. Carollo, D. Cilluffo and F. Ciccarello, Mechanism of decoherence-free coupling between giant atoms, Phys. Rev. Research {\bf 2}, 043184 (2020).

\bibitem{XW2021}X. Wang, T. Liu, A. F. Kockum, H.-R. Li, and F. Nori, Tunable chiral bound states with giant atoms,
 Phys. Rev. Lett. {\bf 126}, 043602 (2021).

\bibitem{XW2022}X. Wang, Z. Gao, J. Li, H. Zhu, and H. Li, Unconventional quantum electrodynamics with a Hofstadter-ladder waveguide, Phys. Rev. A {\bf 106}, 043703 (2022).

\bibitem{AS2021}A. Soro, and A. F. Kockum, Chiral quantum optics with giant atoms, Phys. Rev. A {\bf 105}, 023712 (2022).

\bibitem{NL2022}N. Liu, X. Wang, X. Wang, X. Ma, and M. Cheng, Tunable single photon nonreciprocal scattering based on giant atom-waveguide chiral couplings, Optics Express {\bf 30}, 23428 (2022).


\bibitem{DZ2021}C. A. Gonz\'{a}lez-Guti\'{e}rrez, J. Rom\'{a}n-Roche, and D. Zueco, Distant emitters in ultrastrong waveguide QED: Ground-state properties and non-Markovian dynamics, Phys. Rev. A \textbf{104}, 053701 (2021).


\bibitem{Guoprr20}L. Guo, A. F. Kockum, F. Marquardt, and G. Johansson, Oscillating bound states for a giant atom, Phys. Rev. Research \textbf{2}, 043014 (2020)

\bibitem{SG2020}S. Guo, Y. Wang, T. Purdy, and J. Taylor, Beyond spontaneous emission: Giant atom bounded in the continuum, Phys. Rev. A {\bf 102},
033706 (2020).

\bibitem{Jieqiao} X.-L. Yin, W.-B. Luo, and J.-Q. Liao, Non-Markovian disentanglement dynamics in double-giant-atom waveguide-QED systems, Phys. Rev. A \textbf{106}, 063703 (2022).

\bibitem{PRL2023}A. C. Santos and R. Bachelard, Generation of Maximally Entangled Long-Lived States with Giant Atoms in a Waveguide, Phys. Rev. Lett. {\bf 130}, 053601 (2023).

\bibitem{xian1}L. Du, Y. Zhang, and Y. Li, A giant atom with modulated transition frequency, Front. Phys. {\bf 18}, 12301 (2023).

\bibitem{xian2}S. L. Feng and W. Z. Jia, Manipulating single-photon transport in a waveguide-QED structure containing two giant atoms, Phys. Rev. A {\bf 104}, 063712 (2021).

\bibitem{xian4}Y. Chen, L. Du, L. Guo, Z. Wang, Y. Zhang, Y. Li, and J. Wu, Nonreciprocal and chiral single-photon scattering for giant atoms, Commun. Phys. {\bf 5}, 215 (2022).

\bibitem{c1}M. J. Hartmann, F. G. S. L. Brand\~ao, and M. B. Plenio, Strongly interacting polaritons in coupled arrays of cavities, Nat. Phys. {\bf 2}, 849 (2006).

\bibitem{c2} D. G. Angelakis, M. F. Santos, and S. Bose, Photon-blockade-induced Mott transitions and XY spin models in coupled cavity arrays, Phys. Rev. A {\bf 76}, 031805(R) (2007).

\bibitem{c3}A. A. Houck, H. E. T\"ureci, and J. Koch, On-chip quantum simulation with superconducting circuits, Nat. Phys. {\bf 8}, 292 (2012).

\bibitem{c4}A. Degiron, and D. R. Smith, Nonlinear long-range plasmonic waveguides, Phys. Rev. A {\bf 82}, 033812 (2010).

\bibitem{BIC1}F. H. Stillinger, and D. R. Herrick, Bound states in the continuum, Phys. Rev. A {\bf 11}, 446 (1975).

\bibitem{BIC2}D. C. Marinica, A. G. Borisov, and S. V. Shabanov, Bound States in the Continuum in Photonics, Phys. Rev. Lett. {\bf 100}, 183902 (2008).

\bibitem{BIC3}M. I. Molina, A. E. Miroshnichenko, and Y. S. Kivshar, Surface Bound States in the Continuum, Phys. Rev. Lett. {\bf 108}, 070401 (2012).

\bibitem{BIC4}G. Calaj\'o, Y. L. Fang, H. U. Baranger, and F. Ciccarello, Exciting a Bound State in the Continuum through Multiphoton Scattering Plus Delayed Quantum Feedback, Phys. Rev. Lett. {\bf 122}, 073601 (2019).

\bibitem{BIC5}Q. Qiu, Y. Wu, and X. L\"u, Collective Radiance of Giant Atoms in Non-Markovian Regime, Sci. China Phys. Mech. Astron. {\bf 66}, 224212 (2023).

\bibitem{BOC1}A. Soro, C. S. Mu\~noz, and A. F. Kockum, Interaction between giant atoms in a one-dimensional structured environment, Phys. Rev. A {\bf 107}, 013710 (2023).

\bibitem{BOC2}W. Zhao, and Z. Wang, Single-photon scattering and bound states in an atom-waveguide system with two or multiple coupling points, Phys. Rev. A {\bf 101}, 053855 (2020).

\bibitem{BOC3}Y. Liu, and A. A. Houck, Quantum electrodynamics near a photonic bandgap, Nat. Phys. {\bf 13}, 48 (2016).

\bibitem{BOC4}L. Krinner, M. Stewart, A. Pazmi\~no, J. Kwon, and D. Schneble, Spontaneous emission of matter waves from a tunable open quantum system, Nature {\bf 559}, 589 (2016).

\bibitem{limpra}K. H. Lim, W. Mok, and L. Kwek, Oscillating bound states in non-Markovian photonic lattices, Phys. Rev. A {\bf 107}, 023716 (2023).

\bibitem{mc1}L. Zhou, H. Dong, Y. Liu, C. P. Sun, and F. Nori, Quantum supercavity with atomic mirrors, Phys. Rev. A {\bf 78}, 063827 (2008).

\bibitem{mc2}C. Zhou, Z. Liao, and M. S. Zubairy, Decay of a single photon in a cavity with atomic mirrors, Phys. Rev. A {\bf 105}, 033705 (2022).

\bibitem{mc3}M. Mirhosseini, E. Kim, X. Zhang, A. Sipahigil, P. B. Dieterle, A. J. Keller, A. Asenjo-Garcia, D. E. Chang, and O. Painter, Cavity quantum electrodynamics with atom-like mirrors, Nature {\bf 569}, 692 (2019).

\bibitem{OBC1}{L. Qiao, Y.-J. Song, and C.-P. Sun, Quantum phase transition and interference trapping of populations in a coupled-resonator waveguide, Phys. Rev. A {\bf 100}, 013825 (2021).}

\bibitem{Peter16pra}G. Calaj\'o, F. Ciccarello, D. Chang, and P. Rabl, Atom-field dressed states in slow-light waveguide QED, Phys. Rev. A {\bf 93}, 033833 (2016).

\bibitem{IBC1}{E. N. Bulgakov and A. F. Sadreev, Bound states in the continuum in photonic waveguides inspired by defects, Phys. Rev. B {\bf 78}, 075015 (2008).}



\bibitem{wei}H. Yu, Z. Wang, and J. Wu, Entanglement preparation and nonreciprocal excitation evolution in giant atoms by controllable dissipation and coupling, Phys. Rev. A {\bf 104}, 013720 (2021).

\bibitem{SL}S. Longhi, Rabi oscillations of bound states in the continuum, Opt. Lett. {\bf 46}, 2091 (2021).

\bibitem{An1}C. Chen, C. Yang, and J. An, Exact decoherence-free state of two distant quantum systems in a non-Markovian environment, Phys. Rev. A {\bf 93}, 062122 (2016).

\bibitem{xi1} P. Roushan et al., Spectroscopic signatures of localization with interacting photons in superconducting qubits, Science {\bf 358}, 1175 (2017).

\bibitem{xi2} R. Ma, B. Saxberg, C. Owens, N. Leung, Y. Lu, J. Simon, and D. I. Schuster, A dissipatively stabilized Mott insulator of photons, Nature(London) {\bf 566},51 (2019).

\bibitem{xi3}S. Hacohen-Gourgy, V. V. Ramasesh, C. De Grandi, I. Siddiqi, and S. M. Girvin, Cooling and Autonomous Feedback in a Bose-Hubbard Chain with Attractive Interactions, Phys. Rev. Lett. {\bf 115}, 240501 (2015).

\bibitem{sa} M. Leib, F. Deppe, A. Marx, R. Gross, and M. J. Hartmann, Networks of nonlinear superconducting transmission line resonators, New J. Phys. {\bf 14}, 075024 (2012).

\bibitem{An2}K. Bai, Z. Peng, H. Luo, and J. An, Retrieving Ideal Precision in Noisy Quantum Optical Metrology, Phys. Rev. Lett. {\bf 123}, 040402 (2019).

\bibitem{An3}W. Wu , S. Bai, and J. An, Non-Markovian sensing of a quantum reservoir, Phys. Rev. A {\bf 103}, L010601 (2021).













\end{thebibliography}
\end{document}